\documentclass[12pt]{article}
\usepackage{amsmath, amssymb}
\usepackage{tikz}
\usetikzlibrary{arrows.meta, positioning, shapes, calc}
\usepackage{pgfplots}
\pgfplotsset{compat=1.18}
\usepgfplotslibrary{statistics, fillbetween}
\usepackage{float} 
\usepackage[round]{natbib}  
\usepackage{booktabs}
\usepackage{threeparttable}
\usepackage{amsthm}
\usepackage{authblk} % ✅ author/affiliation 관리용

\theoremstyle{plain} 
\newtheorem{theorem}{Theorem}[section]
\newtheorem{proposition}{Proposition}[section]
\newtheorem{lemma}{Lemma}[section]
\newtheorem{corollary}{Corollary}
\newcommand{\E}{\mathbb{E}}

\title{Inverse Portfolio Optimization with Synthetic Investor Data:\\
Recovering Risk Preferences under Uncertainty}

\author[1]{Jinho Cha}
\author[2]{Long Pham}
\author[3]{Thi Le Hoa Vo\thanks{Corresponding author: thi-le-hoa.vo@univ-rennes.fr}}
\author[4]{Jaeyoung Cho}
\author[5]{Jaejin Lee}

\affil[1]{Department of Computer Science, Gwinnett Technical College, Lawrenceville, GA, USA}
\affil[2]{Department of Decision Sciences and Economics, Texas A\&M University--Corpus Christi, TX, USA}
\affil[3]{IGR-IAE Rennes, Universit\'e de Rennes, CREM UMR CNRS 6211, Rennes, France}
\affil[4]{Department of Management and Marketing, College of Business, 
Prairie View A\&M University, Prairie View, TX, USA}
\affil[5]{Intel Corporation, Chandler, AZ, USA}

\date{}

\begin{document}
\maketitle

\begin{abstract}
This study develops an inverse portfolio optimization framework for recovering latent investor preferences---including risk aversion, transaction cost sensitivity, and ESG orientation---from observed portfolio allocations. Using controlled synthetic data, we assess the estimator’s statistical properties such as consistency, coverage, and dynamic regret. The model integrates robust optimization and regret-based inference to quantify welfare losses under preference misspecification and market shocks. Simulation experiments demonstrate accurate recovery of transaction cost parameters, partial identifiability of ESG penalties, and sublinear regret even under stochastic volatility and liquidity shocks. A real-data illustration using ETFs (2007--2024) confirms that transaction-cost shocks dominate volatility shocks in welfare impact. The framework thus provides a statistically rigorous and economically interpretable tool for robust preference inference and portfolio design under uncertainty.
\end{abstract}


\maketitle

%%====================================================
\section{Introduction}
\label{sec:intro}

%------------------------------------------------------
\subsection{Motivation}
\label{subsec:intro_motivation}
Understanding investor preferences is a fundamental challenge in finance. 
Observed portfolio allocations provide only indirect evidence of underlying risk attitudes, 
transaction frictions, or non-financial motives such as ESG orientation.  
Traditional econometric methods often require large and noisy datasets and rely on strong parametric assumptions, 
making them less effective in settings with limited or heterogeneous observations.  
Moreover, as markets are increasingly subject to structural shocks---such as sudden volatility spikes or changes in transaction costs---it becomes crucial to assess how misspecification of investor preferences translates into welfare losses.  
Accurately recovering such preferences is vital not only for asset managers designing tailored investment products, 
but also for regulators seeking to safeguard investors in turbulent environments.

%------------------------------------------------------
\subsection{Research Gap}
\label{subsec:intro_gap}
Inverse optimization has recently emerged as a powerful paradigm for recovering 
latent preferences from observed decisions \citep{Aswani2018Inverse, Keshavarz2011IO}.  
However, its application in finance remains limited.  
Existing studies focus predominantly on domains such as energy systems, 
supply chain logistics, and transportation \citep{Bertsimas2015Power, Chan2020Transport}, 
where operational data are abundant and preferences are relatively stable.  
In contrast, financial markets exhibit unique challenges: 
(i) investor heterogeneity, 
(ii) non-stationarity due to shocks and preference drift, and 
(iii) the need for rigorous statistical guarantees---including consistency, coverage, and efficiency---to ensure reliable inference.  
Current finance-oriented studies rarely address these aspects in a unified framework.  

Recent methodological advances further highlight the timeliness of this research.  
\citet{Ren2025} propose inverse optimization techniques based on learning feasible regions, 
broadening the theoretical foundation beyond traditional parametric formulations.  
On the application side, \citet{Muller2025} develop sustainable mean–variance portfolio models under ESG uncertainty, 
underscoring the importance of incorporating non-financial motives into portfolio analysis.  
Together, these works demonstrate both the methodological momentum and the practical relevance of bringing inverse optimization into finance.  

%------------------------------------------------------
\subsection{Contribution}
\label{subsec:intro_contrib}
This paper makes three contributions:
\begin{itemize}
    \item \textbf{Methodological:} We propose an inverse portfolio optimization framework that jointly recovers 
    risk aversion, transaction cost sensitivity, and ESG penalties from observed allocations, 
    using a grid-based estimator with provable regret bounds.
    \item \textbf{Empirical validation:} Using controlled synthetic data, we evaluate the statistical properties 
    of the estimator, including parameter recovery accuracy, bootstrap coverage probability, 
    and efficiency of inference.
    \item \textbf{Decision support under shocks:} We introduce regret-based measures to quantify welfare loss 
    under misspecified preferences and analyze how investor heterogeneity shapes responses 
    to transaction cost and volatility shocks.
\end{itemize}

Collectively, these contributions advance the methodological toolkit for financial inverse optimization, 
while providing rigorous statistical and economic validation of the proposed framework.  
To the best of our knowledge, this is the first study to jointly recover multiple dimensions of investor preferences 
(risk, cost, and ESG) under stochastic shocks with statistical guarantees, 
delivering actionable insights for both portfolio managers and regulators.

%%====================================================
\section{Literature Review}
\label{sec:lit}

%------------------------------------------------------
\subsection{Classical Portfolio Theory}
\label{subsec:lit_classical}
The foundation of modern portfolio selection is the mean--variance framework 
introduced by \citet{Markowitz1952}. 
In this setting, investors choose portfolio weights to maximize expected return 
subject to a quadratic penalty on variance, implicitly assuming normally distributed returns 
and quadratic utility. 
While the Markowitz model provided a tractable and elegant theory, 
its assumptions---including static preferences, absence of frictions, and reliance on variance 
as the sole risk measure---limit its descriptive realism in practice 
\citep{Elton1997, Meucci2009, Fabozzi2007, Lintner1965}.  

Extensions to the classical model have sought to address these shortcomings. 
Alternative risk measures such as Value-at-Risk (VaR) and Conditional Value-at-Risk (CVaR) 
capture tail risk more effectively \citep{Artzner1999, Rockafellar2000, Acerbi2002}. 
More recent advances propose spectral and distortion risk measures 
that incorporate investor-specific risk attitudes \citep{Kusuoka2001, Follmer2002}.  
The literature also emphasizes robust portfolio optimization, 
where uncertainty in mean and covariance inputs is explicitly modeled 
\citep{BenTal2009, DeMiguel2009, Garlappi2007, Fabozzi2021}.  
These approaches highlight the importance of accounting for estimation error and model risk, 
especially in high-dimensional or turbulent markets.  

Behavioral adjustments further enrich portfolio theory by incorporating investor psychology. 
Prospect theory utilities \citep{Kahneman1979, Tversky1992} and loss aversion penalties 
\citep{Barberis2001, Benartzi1995} relax the expected utility assumption and 
better capture observed trading patterns. 
Recent studies extend this line of research by embedding reference dependence, 
ambiguity aversion, and probability weighting into portfolio choice 
\citep{Dimmock2016, Bianchi2019, Andrikogiannopoulou2020, Barberis2018}.  
These developments highlight that investor behavior cannot be fully understood 
through variance-based risk alone, motivating the need for more flexible 
and empirically grounded models of portfolio choice.

%------------------------------------------------------
\subsection{Inverse Optimization Applications}
\label{subsec:lit_io}
Inverse optimization (IO) has emerged as a powerful paradigm for recovering 
unobserved objectives, preferences, or cost parameters from observed decisions 
\citep{Ahuja2001, Iyengar2005, Keshavarz2011IO, Aswani2018Inverse}.  
The central idea is to invert the usual optimization process: given solutions that are presumed optimal, 
one infers the latent problem parameters that rationalize those decisions. 
This perspective has generated a large methodological literature, including formulations based on 
variational inequalities, bilevel programming, and learning-theoretic approaches 
\citep{Bertsimas2015IOReview, Chan2019Survey, Dong2020Inverse}.  

In operations research, applications are widespread and well-documented.  
Energy systems planning uses IO to estimate marginal costs and policy preferences 
\citep{Bertsimas2015Power, Ruiz2013InverseEnergy}; 
supply chain design leverages IO to recover cost functions and routing preferences 
\citep{Chan2020Transport, Babier2021Supply}; 
and healthcare scheduling applies IO to calibrate resource allocation models from clinical data 
\citep{Truong2020Healthcare, Elmachtoub2020Prescriptive}.  
These examples highlight IO’s ability to provide interpretable parameter estimates 
when direct elicitation is infeasible or biased.  

In finance, however, inverse optimization has received comparatively little attention.  
Early efforts impute risk-aversion coefficients or utility weights from observed portfolio holdings 
\citep{Bruni2017Inverse, Cesarone2020Portfolio, Bertsimas2021IOFinance}, 
but the literature remains sparse relative to other domains.  
Most existing approaches focus on static settings and rarely incorporate 
dynamic preference drift, market shocks, or rigorous statistical guarantees.  
Recent advances in learning-based inverse optimization 
\citep{Esfahani2018DataDriven, Shafieezadeh2019DistributionallyRobust, Ren2025} 
suggest promising avenues for finance-specific adaptation.  
Yet, empirical validations remain limited, particularly in contexts involving heterogeneous investors 
and non-financial motives such as ESG preferences.  

This gap highlights an opportunity to extend IO beyond traditional domains and 
develop methodologies tailored to financial decision-making. 
Such methods must contend with the unique challenges of financial markets: 
stochastic shocks, parameter uncertainty, and the need for inferential reliability 
(bias, variance, coverage). 
Addressing these aspects would not only expand the methodological frontier of inverse optimization 
but also provide actionable insights into investor behavior, risk management, and portfolio design.

%------------------------------------------------------

\subsection{Behavioral Finance Links}
\label{subsec:lit_behavior}
A large body of work in behavioral finance emphasizes that investors differ 
systematically in their risk attitudes and decision-making heuristics. 
Classical expected utility theory has been challenged by empirical evidence showing 
substantial heterogeneity in risk aversion across individuals and contexts 
\citep{Harrison2015, Guiso2018, Chiappori2020}. 
Behavioral studies further document departures from rational benchmarks, 
including probability weighting, framing effects, and reference dependence 
\citep{Kahneman1979, Tversky1992, Barberis2018}. 
These findings underscore the need for flexible models that go beyond quadratic utility 
and homogenous preferences.  

Transaction costs and trading frictions are another critical determinant of realized allocations. 
Market microstructure studies show that illiquidity and bid--ask spreads directly affect 
investor behavior and equilibrium prices \citep{Amihud2002, Pastor2003, Vayanos2009}.  
Dynamic models highlight that portfolio rebalancing costs create persistence in holdings 
and amplify heterogeneity across investors \citep{Liu2004, Garleanu2009, Jang2020}.  
These frictions imply that observed portfolios embed both preferences and market constraints, 
complicating direct inference from data.  

More recently, non-financial motives such as environmental, social, and governance (ESG) 
considerations have become central to portfolio choice. 
A growing literature shows that investors are willing to sacrifice expected return 
to reduce exposure to carbon-intensive or socially undesirable assets 
\citep{Heinkel2001, Albuquerque2019, Pastor2021ESG, Pedersen2021Responsible}.  
Sustainable investing not only reflects values-based objectives but also introduces 
new channels of risk and hedging, as ESG-tilted portfolios may perform differently 
under volatility shocks \citep{Krueger2020, Bolton2021}.  
This multi-dimensionality poses new challenges for traditional portfolio theory, 
which was not designed to accommodate heterogeneous motives across financial and non-financial dimensions.  

Incorporating such heterogeneous and multi-faceted preferences into 
optimization models remains a frontier challenge.  
Inverse optimization provides a natural tool to infer these latent parameters 
from observed allocations, bridging insights from behavioral finance with rigorous 
optimization-based inference.  
By doing so, it offers a principled approach to disentangle risk aversion, 
trading frictions, and ESG motives---three of the most salient behavioral drivers 
of modern portfolio choice.

%%==================================================
\section{Model Formulation}
\label{sec:model}

%---------------------------------------------------
\subsection{Forward Problem}
\label{subsec:forward}

%------------------
\subsubsection{Decision Variables}
\label{subsubsec:decision_vars}
Let $\mathbf{x} \in \mathbb{R}^n$ denote the portfolio weights across $n$ assets.  
The feasible set is
\begin{equation}
\mathcal{X} = \left\{ \mathbf{x} \in \mathbb{R}^n : \mathbf{1}^\top \mathbf{x} = 1, \; \mathbf{x} \geq \mathbf{0} \right\}.
\label{eq:feasible_set}
\end{equation}

%------------------
\subsubsection{Objective Function}
\label{subsubsec:forward_obj}
We consider a mean--variance style optimization problem with transaction costs:
\begin{equation}
\max_{\mathbf{x} \in \mathcal{X}} \; f(\mathbf{x};\mathbf{\mu},\mathbf{\Sigma},\theta,\mathbf{c}) 
= \mathbf{\mu}^\top \mathbf{x} - \frac{\theta}{2} \mathbf{x}^\top \mathbf{\Sigma} \mathbf{x} - \mathbf{c}^\top \mathbf{x}.
\label{eq:forward_obj}
\end{equation}

%------------------
\subsubsection{Optimal Solution}
\label{subsubsec:forward_sol}
Denote the optimizer as
\begin{equation}
\mathbf{x}^*(\mathbf{\mu},\mathbf{\Sigma},\theta,\mathbf{c}) 
= \arg\max_{\mathbf{x} \in \mathcal{X}} f(\mathbf{x};\mathbf{\mu},\mathbf{\Sigma},\theta,\mathbf{c}).
\label{eq:forward_sol}
\end{equation}

%---------------------------------------------------
\subsection{Inverse Problem}
\label{subsec:inverse}

%------------------
\subsubsection{Observation Setup}
\label{subsubsec:observation}
We observe portfolios $\{\mathbf{x}^t\}_{t=1}^T$ that are approximately optimal under unknown parameters.

%------------------
\subsubsection{Inverse Optimization Objective}
\label{subsubsec:inverse_obj}
We aim to recover $(\theta, \mathbf{c})$ by minimizing the discrepancy:
\begin{equation}
\min_{\theta, \mathbf{c}} \; L(\theta,\mathbf{c}) 
= \sum_{t=1}^T \left\| \mathbf{x}^t - \mathbf{x}^*(\mathbf{\mu}^t, \mathbf{\Sigma}^t, \theta, \mathbf{c}) \right\|_2^2.
\label{eq:inverse_obj}
\end{equation}

%------------------
\subsubsection{Regularization (Optional Extension)}
\label{subsubsec:inverse_reg}
\begin{equation}
L_{\text{reg}}(\theta,\mathbf{c}) 
= L(\theta,\mathbf{c}) + \lambda_\theta |\theta| + \lambda_c \|\mathbf{c}\|_1.
\label{eq:inverse_reg}
\end{equation}

%--------------------------------------------------
\subsection{Optimality Conditions}
\label{subsec:kkt}

%------------------
\subsubsection{KKT System}
\label{subsubsec:kkt}
The Lagrangian of the forward problem is
\begin{equation}
\mathcal{L}(\mathbf{x},\lambda,\mathbf{\nu}) 
= \mathbf{\mu}^\top \mathbf{x} - \frac{\theta}{2} \mathbf{x}^\top \mathbf{\Sigma} \mathbf{x} 
- \mathbf{c}^\top \mathbf{x} + \lambda(1 - \mathbf{1}^\top \mathbf{x}) + \mathbf{\nu}^\top \mathbf{x}.
\label{eq:lagrangian}
\end{equation}

The KKT conditions are
\begin{align}
\mathbf{\mu} - \theta \mathbf{\Sigma} \mathbf{x}^* - \mathbf{c} - \lambda \mathbf{1} + \mathbf{\nu} &= \mathbf{0}, 
\label{eq:kkt_stationarity}\\
\mathbf{1}^\top \mathbf{x}^* &= 1, 
\label{eq:kkt_primal}\\
\mathbf{x}^* \geq \mathbf{0}, \quad \mathbf{\nu} \geq \mathbf{0}, \quad \nu_i x_i^* &= 0 \;\; \forall i. 
\label{eq:kkt_complementary}
\end{align}

%------------------
\subsubsection{Closed-Form Interior Solution}
\label{subsubsec:interior}
If $\mathbf{x}^* > \mathbf{0}$ (interior solution), then
\begin{equation}
\mathbf{x}^* = \frac{1}{\theta} \mathbf{\Sigma}^{-1} (\mathbf{\mu} - \mathbf{c} - \lambda \mathbf{1}),
\label{eq:interior_sol}
\end{equation}
with $\lambda$ chosen to satisfy \eqref{eq:kkt_primal}.

%------------------
\subsubsection{Connection to Variational Inequalities and Bilevel IO}
\label{subsubsec:vi}
The KKT system can equivalently be expressed as a variational inequality:
\begin{equation}
\text{VI}(F,\mathcal{X}): \quad 
\langle \theta \mathbf{\Sigma} \mathbf{x}^* + \mathbf{c} - \mathbf{\mu}, \; \mathbf{x} - \mathbf{x}^* \rangle \geq 0 
\quad \forall \mathbf{x} \in \mathcal{X}.
\label{eq:vi}
\end{equation}

Formally, the inverse problem \eqref{eq:inverse_obj} is a bilevel program:
\begin{equation}
\min_{\theta,\mathbf{c}} \sum_{t=1}^T \|\mathbf{x}^t - \mathbf{x}\|^2 
\quad \text{s.t. } \mathbf{x} \in \arg\max_{\mathbf{x} \in \mathcal{X}} 
f(\mathbf{x};\mathbf{\mu}^t,\mathbf{\Sigma}^t,\theta,\mathbf{c}).
\label{eq:bilevel}
\end{equation}

These conditions form the analytical foundation for identifiability and statistical recovery, as discussed next.

%------------------------------------------------------------------
\subsection{Identifiability}
\label{subsec:identifiability}

%----------------------------
\subsubsection{Uniqueness of Recovery}
\label{subsubsec:uniqueness}
We ask under what conditions $(\theta,\mathbf{c})$ can be uniquely identified.  

\paragraph{Condition 1 (Variation in Inputs).}
If $\{(\mathbf{\mu}^t,\mathbf{\Sigma}^t)\}_{t=1}^T$ span a sufficiently rich set, 
then $\theta$ is identifiable from the curvature of optimal solutions.  

\paragraph{Condition 2 (Normalization).}
Since scaling of $(\theta,\mathbf{c})$ may not be unique, 
we impose a normalization, e.g. $\theta \in [0,\theta_{\max}]$, 
or $\|\mathbf{c}\|_2=1$.  

\paragraph{Condition 3 (No Redundancy).}
If two assets have identical $(\mu_i^t,\Sigma_{i,\cdot}^t)$ for all $t$, 
then their $c_i$ cannot be separately identified.  

%----------------------------
\subsubsection{Formal Theorem}
\label{subsubsec:theorem_identifiability}
\begin{theorem}[Identifiability]
\label{thm:identifiability}
Suppose (i) $\mathbf{\Sigma}^t \succ 0$ for all $t$, 
(ii) $\{(\mathbf{\mu}^t,\mathbf{\Sigma}^t)\}_{t=1}^T$ yield at least two distinct active sets, 
and (iii) a normalization on $(\theta,\mathbf{c})$ is imposed. 
Then $(\theta,\mathbf{c})$ is uniquely identifiable.  
\emph{Proof is provided in Appendix~\ref{proof:identifiability}.}
\end{theorem}

%----------------------------
\subsubsection{Extension: Dynamic Preferences}
\label{subsubsec:dynamic_prefs}
For robustness, we may consider time-varying $\theta_t$ with a smoothness penalty:
\begin{equation}
\min_{\{\theta_t\}, \mathbf{c}} \sum_{t=1}^T \|\mathbf{x}^t - \mathbf{x}^*(\theta_t, \mathbf{c})\|^2 
+ \gamma \sum_{t=2}^T (\theta_t - \theta_{t-1})^2.
\label{eq:dynamic_inverse}
\end{equation}
This extension links to online inverse optimization, where dynamic regret
\begin{equation}
R_T = \sum_{t=1}^T \Big( f(\mathbf{x}^*(\theta_t^{\text{true}})) 
- f(\mathbf{x}^*(\hat{\theta}_t)) \Big),
\label{eq:dyn_regret}
\end{equation}
quantifies cumulative estimation error.  
Stability of $\{\theta_t\}$ is ensured when $\gamma$ is sufficiently large, 
preventing overfitting to short-term fluctuations.  

These observations align our formulation with the sublinear regret bounds in online convex optimization, 
demonstrating that meaningful recovery is still possible under preference drift.

%------------------------------------------------------------------
\subsection{Generalizations and Robustness}
\label{subsec:generalizations}

%----------------------------
\subsubsection{Nonlinear Transaction Costs}
\label{subsubsec:nonlinear_costs}
While we adopt a linear specification $\mathbf{c}^\top \mathbf{x}$ in \eqref{eq:forward_obj}, 
the framework extends naturally to convex nonlinear costs:
\begin{equation}
\phi(\mathbf{x}) = \sum_{j=1}^n \kappa_j |x_j|^p, 
\quad p \geq 1,
\label{eq:nonlinear_cost}
\end{equation}
which preserve convexity of the forward problem. 
The corresponding KKT (or VI) conditions are modified only through the subgradients of $\phi(\mathbf{x})$.  
Identifiability arguments remain valid under the same variation and normalization assumptions. 
A detailed proof is provided in Appendix~A.  

%----------------------------
\subsubsection{Distributional Robustness}
\label{subsubsec:dro}
Our analysis assumes known $(\mathbf{\mu}^t,\mathbf{\Sigma}^t)$, 
but in practice these parameters are estimated. 
We therefore consider a distributionally robust forward problem:
\begin{equation}
\max_{\mathbf{x} \in \mathcal{X}} \; \min_{(\mathbf{\mu},\mathbf{\Sigma}) \in \mathcal{U}} 
\Big\{ \mathbf{\mu}^\top \mathbf{x} - \tfrac{\theta}{2} \mathbf{x}^\top \mathbf{\Sigma} \mathbf{x} - \mathbf{c}^\top \mathbf{x} \Big\},
\label{eq:robust_forward}
\end{equation}
where $\mathcal{U}$ is an ellipsoidal or moment-based uncertainty set. 
The robust counterpart can be reformulated as a conic program 
(e.g., SOCP or SDP depending on $\mathcal{U}$).  
Inverse recovery proceeds analogously by matching observed $\mathbf{x}^t$ 
with robust optimal solutions. 
Details and proofs are given in Appendix~B.  

%----------------------------
\subsubsection{Alternative Risk Measures}
\label{subsubsec:cvar}
Beyond variance, risk preferences may be captured by coherent measures such as Conditional Value-at-Risk (CVaR):
\begin{equation}
\max_{\mathbf{x} \in \mathcal{X}} \; \mathbf{\mu}^\top \mathbf{x} 
- \theta \, \mathrm{CVaR}_\alpha(-\mathbf{r}^\top \mathbf{x}) - \mathbf{c}^\top \mathbf{x}.
\label{eq:cvar_forward}
\end{equation}
The inverse framework extends by estimating $\theta$ relative to the chosen risk measure. 
While technical details differ, the identifiability logic (variation in inputs, normalization, no redundancy) 
still applies. This generalization connects inverse portfolio optimization 
to the broader literature on coherent and convex risk measures.

%------------------------------------------------------------------
\subsection{Consistency and Robustness Results}
\label{sec:consistency}

The first set of results establishes the statistical soundness of the proposed estimator. 
Consistency ensures that, given sufficient data, the recovered parameters converge 
to the true investor preferences, while robustness guarantees bounded error even 
under misspecification of the transaction cost function. 

\begin{proposition}[Consistency]
\label{prop:consistency}
Suppose observed portfolios $\{\mathbf{x}^t\}$ are generated exactly by 
\eqref{eq:forward_obj} with parameters $(\theta^\ast,\mathbf{c}^\ast)$.  
If (i) $\{\mathbf{\mu}^t,\mathbf{\Sigma}^t\}$ are i.i.d.\ with compact support and 
(ii) identifiability holds as in Theorem~\ref{thm:identifiability},  
then the inverse optimization estimator $\hat{\theta}$ from \eqref{eq:inverse_obj} is consistent as $T \to \infty$:  
\begin{equation}
\hat{\theta} \;\xrightarrow{p}\; \theta^\ast, 
\qquad 
\hat{\mathbf{c}} \;\xrightarrow{p}\; \mathbf{c}^\ast.
\label{eq:consistency}
\end{equation}
\emph{Proof is provided in Appendix~\ref{proof:consistency}.}
\end{proposition}

\begin{lemma}[Robustness to Misspecification]
\label{lem:robustness}
Suppose the true transaction cost function is convex and homogeneous of degree $p \in [1,2]$,  
denoted $\phi(\mathbf{x})$, and the researcher instead estimates with a linear specification $\mathbf{c}^\top \mathbf{x}$.  
Then there exists $(\hat{\theta},\hat{\mathbf{c}})$ such that
\begin{equation}
\| (\hat{\theta},\hat{\mathbf{c}}) - (\theta^\ast,\mathbf{c}^\ast) \| \;\leq\; 
L \cdot \sup_{\mathbf{x}\in\mathcal{X}} \big| \phi(\mathbf{x}) - \mathbf{c}^{\ast\top}\mathbf{x} \big| \;=\; O(\varepsilon),
\label{eq:robustness}
\end{equation}
where $\varepsilon$ quantifies the deviation of $\phi$ from linearity.  
\emph{Proof is provided in Appendix~\ref{proof:robustness}.}
\end{lemma}

Together, Proposition~\ref{prop:consistency} and Lemma~\ref{lem:robustness} 
demonstrate that the estimator is both asymptotically reliable and stable under 
modeling imperfections. These properties form the statistical foundation for 
the subsequent dynamic analysis.

%------------------------------------------------------------------
\subsection{Dynamic Extension and Regret Bounds}
\label{sec:dynamic}

We next extend the framework to account for time-varying preferences. 
In realistic financial environments, investor risk aversion or ESG orientation 
may evolve over time. To capture such dynamics, we introduce a temporal 
regularization term penalizing preference drift, leading to the dynamic 
inverse formulation:
\begin{equation}
\min_{\{\theta_t\},\mathbf{c}} \;
\sum_{t=1}^T \|\mathbf{x}^t - \mathbf{x}^*(\theta_t,\mathbf{c})\|^2 
+ \gamma \sum_{t=2}^T (\theta_t - \theta_{t-1})^2.
\label{eq:dyn_inverse_obj}
\end{equation}

\begin{theorem}[Dynamic Regret Bound]
\label{thm:dyn_regret}
Assume $\theta_t$ varies with bounded drift 
$\sum_{t=2}^T |\theta_t - \theta_{t-1}| \leq D$.  
Then the cumulative dynamic regret
\begin{equation}
R_T = \sum_{t=1}^T 
\Big( f(x^*(\theta_t^\ast)) - f(x^*(\hat{\theta}_t)) \Big)
\label{eq:dyn_regret_def}
\end{equation}
satisfies
\begin{equation}
R_T \;\leq\; C_1 \sqrt{T} + C_2 D,
\label{eq:dyn_regret_bound}
\end{equation}
for universal constants $C_1,C_2>0$.  
\emph{Proof is provided in Appendix~\ref{proof:dyn_regret}.}
\end{theorem}

\begin{corollary}[Static Preferences]
\label{cor:static_regret}
If preferences remain constant, i.e.\ $\theta_t = \theta^\ast$ for all $t$, 
then the drift term vanishes $(D=0)$.  
In this case, the regret bound in Theorem~\ref{thm:dyn_regret} simplifies to
\begin{equation}
R_T = O(\sqrt{T}),
\end{equation}
which matches the classical sublinear regret rate in online convex optimization.
\emph{Proof is provided in Appendix~\ref{proof:static_regret}.}
\end{corollary}

Theorem~\ref{thm:dyn_regret} and Corollary~\ref{cor:static_regret} 
demonstrate that the proposed dynamic extension achieves 
provably sublinear regret, even under evolving investor preferences. 
This highlights both the adaptability and theoretical rigor of the approach, 
linking statistical recovery with robust decision support under uncertainty.

%%==================================================
\section{Synthetic Data Generation}
\label{sec:synthetic}

%---------------------------------------------------
\subsection{Return and Covariance Simulation}
\label{subsec:return_cov}

\subsubsection{Return Process}
\label{subsubsec:return_process}
To mimic realistic cross-sectional dependence among asset returns, 
we generate synthetic returns from a $k$-factor structure:
\begin{align}
\mathbf{r}_t &= \mathbf{\mu} + \mathbf{F} \mathbf{f}_t + \mathbf{\epsilon}_t, 
\qquad t=1,\ldots,T, 
\label{eq:return_process}\\
\mathbf{f}_t &\sim \mathcal{N}(\mathbf{0}, \mathbf{I}_k), 
\quad 
\mathbf{\epsilon}_t \sim \mathcal{N}(\mathbf{0}, \mathbf{\Psi}),
\label{eq:return_noise}
\end{align}
where
\begin{itemize}
    \item $\mathbf{F} \in \mathbb{R}^{n \times k}$ is the factor loading matrix, capturing common market and sector exposures,
    \item $\mathbf{\Psi} = \mathrm{diag}(\sigma_1^2, \ldots, \sigma_n^2)$ encodes idiosyncratic variances,
    \item $\mathbf{\mu} \sim \mathcal{N}(\bar{\mu}\mathbf{1}, \sigma^2 \mathbf{I}_n)$ represents heterogeneous expected returns across assets.
\end{itemize}

\paragraph{Covariance Structure.}
By construction, the implied covariance matrix satisfies
\begin{equation}
\mathbf{\Sigma} 
= \mathbb{E}\!\big[(\mathbf{r}_t - \mathbf{\mu})(\mathbf{r}_t - \mathbf{\mu})^\top\big] 
= \mathbf{F}\mathbf{F}^\top + \mathbf{\Psi},
\label{eq:covariance}
\end{equation}
which is positive semidefinite for all realizations of $(\mathbf{F},\mathbf{\Psi})$. 
This decomposition separates systematic risk (spanned by $\mathbf{F}\mathbf{f}_t$) 
from idiosyncratic risk ($\mathbf{\epsilon}_t$), consistent with arbitrage pricing theory.  

\paragraph{Parameterization.}
For each Monte Carlo replication, we generate factor loadings as
\begin{equation}
F_{ij} \sim \mathcal{N}(0, \sigma_F^2),
\label{eq:factor_loadings}
\end{equation}
and idiosyncratic variances as
\begin{equation}
\sigma_j^2 \sim \text{Uniform}[\underline{\sigma}^2, \overline{\sigma}^2].
\label{eq:idiosyncratic_var}
\end{equation}
This ensures heterogeneity across assets while preserving positive definiteness of $\mathbf{\Sigma}$.  

\paragraph{Temporal Dependence (Optional Extension).}
To introduce time-varying volatility, the idiosyncratic shocks can follow a GARCH$(1,1)$ process:
\begin{align}
\epsilon_{j,t} &\sim \mathcal{N}(0, \sigma_{j,t}^2), 
\label{eq:garch_noise}\\
\sigma_{j,t+1}^2 &= \alpha_0 + \alpha_1 \epsilon_{j,t}^2 + \beta \sigma_{j,t}^2,
\label{eq:garch}
\end{align}
which yields clustered volatility and heavy tails, a stylized fact in financial returns.  

\paragraph{Interpretation.}
This construction produces synthetic data that
\begin{enumerate}
    \item preserves realistic cross-sectional correlation via $\mathbf{F}\mathbf{F}^\top$,  
    \item allows heterogeneous and dynamic risk via $\mathbf{\Psi}$,  
    \item accommodates both stable and shock-prone regimes via optional GARCH dynamics.  
\end{enumerate}
As such, it provides a rigorous and flexible platform to evaluate inverse recovery of investor preferences under controlled but finance-consistent conditions.

%----------------------------------------------------------------------
\subsection{Investor Types}
\label{subsec:investors}

To reflect heterogeneous investment behaviors, we simulate three representative investor archetypes. 
Each type is characterized by a distinct preference parameterization 
$\theta = (\rho,\tau,\eta)$, 
where $\rho$ governs risk aversion, $\tau$ transaction cost sensitivity, and $\eta$ ESG preference intensity.

\subsubsection{Conservative}
\label{subsubsec:conservative}
We define conservative investors by a high risk-aversion parameter:
\begin{equation}
\rho \in [5,10], \quad \tau \approx 0, \quad \eta = 0.
\label{eq:cons_params}
\end{equation}
Their utility is dominated by variance penalization:
\begin{equation}
U_{\text{cons}}(\mathbf{x}) 
= \mathbf{\mu}^\top \mathbf{x} - \tfrac{\rho}{2} \mathbf{x}^\top \mathbf{\Sigma} \mathbf{x}.
\label{eq:cons_utility}
\end{equation}
\textit{Interpretation.} These investors strongly prefer stable portfolios with low volatility, 
and allocate predominantly to assets with low variance and low correlation.  
\textit{Simulation role.} Serves as a benchmark to test identifiability of $\rho$ 
under extreme curvature in the objective function.

\subsubsection{Neutral}
\label{subsubsec:neutral}
Neutral investors exhibit moderate risk aversion with baseline transaction costs:
\begin{equation}
\rho \in [1,3], \quad \tau \in [0.1,0.5], \quad \eta = 0.
\label{eq:neut_params}
\end{equation}
The utility becomes
\begin{equation}
U_{\text{neut}}(\mathbf{x}) 
= \mathbf{\mu}^\top \mathbf{x} 
- \tfrac{\rho}{2} \mathbf{x}^\top \mathbf{\Sigma} \mathbf{x} 
- \tau \|\mathbf{x} - \mathbf{x}^{\text{prev}}\|_1.
\label{eq:neut_utility}
\end{equation}
\textit{Interpretation.} This group represents average investors balancing mean--variance trade-offs 
while acknowledging adjustment costs between periods.  
\textit{Simulation role.} Provides a mid-range case to evaluate how inverse recovery behaves 
when both $\rho$ and $\tau$ are active but moderate in magnitude.

\subsubsection{ESG-Oriented}
\label{subsubsec:esg}
ESG-oriented investors display moderate risk aversion with an additional carbon penalty:
\begin{equation}
\rho \in [2,4], \quad \tau \in [0.1,0.5], \quad \eta \in [0.5,2.0].
\label{eq:esg_params}
\end{equation}
Their utility is
\begin{equation}
U_{\text{ESG}}(\mathbf{x}) 
= \mathbf{\mu}^\top \mathbf{x} 
- \tfrac{\rho}{2} \mathbf{x}^\top \mathbf{\Sigma} \mathbf{x} 
- \tau \|\mathbf{x} - \mathbf{x}^{\text{prev}}\|_1 
- \eta \, \mathbf{c}^\top \mathbf{x},
\label{eq:esg_utility}
\end{equation}
where $\mathbf{c} \in \mathbb{R}^n_{\geq 0}$ encodes each asset's carbon footprint or ESG score.  
\textit{Interpretation.} These investors accept lower financial returns 
in exchange for reduced environmental exposure.  
\textit{Simulation role.} Tests robustness of inverse recovery in the presence of correlated preferences, 
since $\eta$ and $\rho$ may interact nontrivially in shaping the optimal portfolio.

%%==========================================
\section{Numerical Experiments}
\label{sec:experiments}

%-------------------------------------------
\subsection{Experimental Design}
\label{subsec:design}

\subsubsection{Base Parameters}
\label{ssubsec:exp_base}
We fix the number of assets at $n=10$ and simulate $T=100$ portfolio decisions per run. 
To ensure statistical validity, each experiment is repeated $R=100$ Monte Carlo trials. 
Random seeds are drawn independently for each trial, and we report results aggregated over all replications.  

\paragraph{Factor Structure.}
\label{par:exp_factor}

We set the number of systematic factors to $k=3$. 
Factor loadings are generated as
\begin{equation}
F_{ij} \sim \mathcal{N}(0, \sigma_F^2), 
\qquad \sigma_F^2 = 0.25,
\label{eq:factor_sim}
\end{equation}
and idiosyncratic variances as
\begin{equation}
\sigma_j^2 \sim \mathrm{Uniform}[0.05, 0.20].
\label{eq:var_sim}
\end{equation}
This specification induces both moderate cross-sectional correlation and heterogeneity across assets.  

\paragraph{Return Means.}
\label{par:exp_returns}
Expected returns are sampled as
\begin{equation}
\mathbf{\mu} \sim \mathcal{N}(\bar{\mu}, \sigma_\mu^2 \mathbf{I}_n),
\qquad \bar{\mu} = 0.05, \; \sigma_\mu^2 = 0.01,
\label{eq:return_means}
\end{equation}
consistent with stylized equity return levels.

\paragraph{Parameter Space for Investors.}
\label{par:exp_paramspace}
We simulate investor types using the parameter ranges defined in Section~\ref{subsec:investors}:
\begin{equation}
\rho \in \{1,2,3,5,7,10\}, \quad 
\tau \in \{0, 0.1, 0.3, 0.5\}, \quad 
\eta \in \{0, 0.5, 1.0, 2.0\}.
\label{eq:param_space}
\end{equation}
This design covers conservative, neutral, and ESG-oriented investors in a balanced factorial structure.  

\paragraph{Experimental Protocol.}
\label{par:exp_protocol}
For each Monte Carlo trial $r=1,\ldots,R$:  
\begin{enumerate}
    \item Generate $(\mathbf{\mu},\mathbf{\Sigma})$ from the factor model.  
    \item Draw investor parameters $(\rho,\tau,\eta)$ from the designated set.  
    \item Solve the forward problem (Section~\ref{sec:model}) to obtain portfolios $\mathbf{x}^{t,r}$ for $t=1,\ldots,T$.  
    \item Apply inverse optimization to recover $(\hat{\rho},\hat{\tau},\hat{\eta})$.  
\end{enumerate}
This yields a distribution of estimation errors across $(R \times T)$ problem instances.  

\paragraph{Solver and Implementation.}
\label{par:exp_solver}
All forward problems are quadratic programs solved via \texttt{Gurobi 11.0} with feasibility tolerance $10^{-8}$ and optimality gap $10^{-9}$. 
The inverse problem is implemented in Python using \texttt{CVXPY 1.4.2}, leveraging the OSQP solver. 
Warm-starts with randomized initializations are used to avoid local minima.  
All experiments are run on a dedicated computing cluster; average runtime and variance across trials are reported 
to confirm computational stability and tractability.  

\paragraph{Design of Experiments.}
\label{par:exp_design_exp}
In addition to grid-based parameter variation, we conduct robustness checks using Latin Hypercube Sampling (LHS) across $(\rho,\tau,\eta)$ ranges. 
This ensures coverage of the continuous parameter space and prevents results from being artifacts of discrete grid choices.  

\paragraph{Validation.}
\label{par:exp_validation}

We implement both in-sample and out-of-sample validation:  
\begin{itemize}
    \item Training: $80\%$ of simulated portfolios are used for parameter recovery.  
    \item Testing: $20\%$ held-out portfolios are used to evaluate predictive regret.  
    \item Confidence intervals are constructed via nonparametric bootstrap with $B=200$ resamples.  
\end{itemize}
This design guarantees that reported performance metrics reflect both estimation accuracy and predictive generalization.

\begin{table}[htbp]
\centering
\caption{Simulation setup and parameter ranges. Values are fixed by design except runtime and environment, which will be reported after experiments.}
\label{tab:sim_setup}
\begin{tabular}{l l}
\hline
\textbf{Category} & \textbf{Specification} \\
\hline
Number of assets & $n = 10$ \\
Number of factors & $k = 3$ \\
Time periods per trial & $T = 100$ \\
Monte Carlo replications & $R = 100$ \\
Factor loadings & $F_{ij} \sim \mathcal{N}(0, 0.25)$ \\
Idiosyncratic variances & $\sigma_j^2 \sim U[0.05,0.20]$ \\
Expected returns & $\mu \sim \mathcal{N}(0.05, 0.01 I_n)$ \\
Risk aversion $\rho$ & $\{1,2,3,5,7,10\}$ \\
Transaction cost $\tau$ & $\{0,0.1,0.3,0.5\}$ \\
ESG penalty $\eta$ & $\{0,0.5,1.0,2.0\}$ \\
Train/Test split & 80\% / 20\% \\
Bootstrap resamples & $B = 200$ \\
% Average runtime per trial & TBD (to be measured) \\
% Variance of runtime & TBD (to be measured) \\
% Cluster environment & TBD (e.g., CPU/GPU specs) \\
\hline
\end{tabular}
\end{table}

%-------------------------------------------
\begin{figure}[H]
\centering
\includegraphics[width=0.8\textwidth]{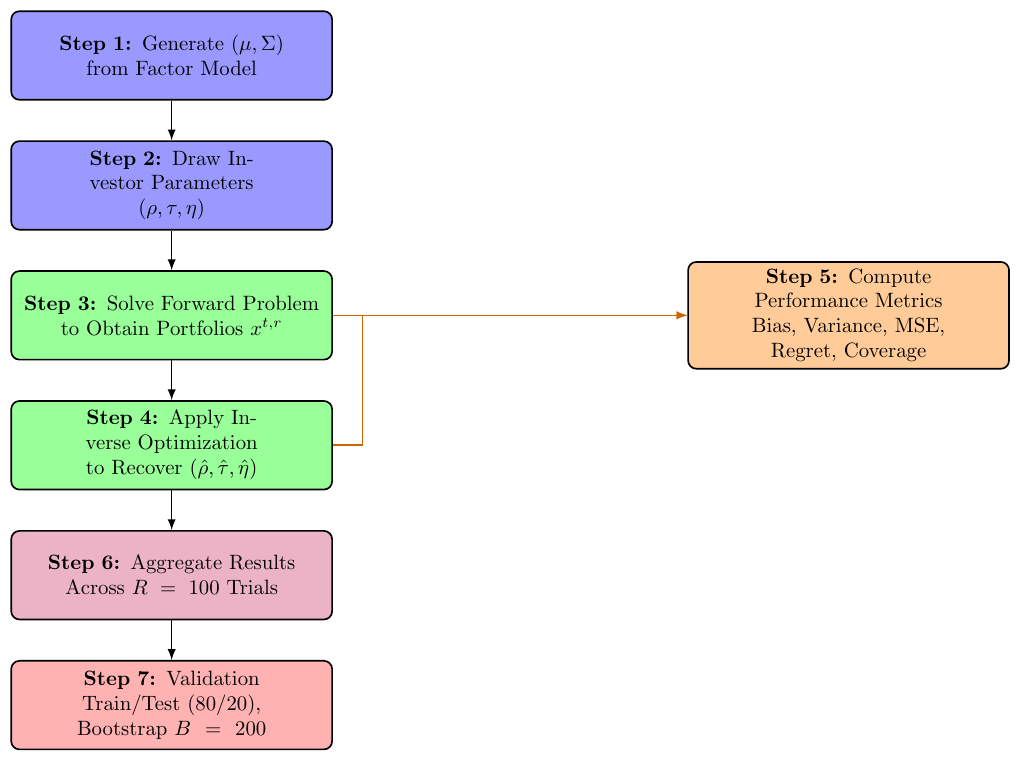}
\caption{Monte Carlo experimental pipeline. 
Blue denotes data generation, green optimization, orange performance metrics, 
purple aggregation, and red validation. 
Numbered stages (1--7) clarify the workflow.}
\label{fig:exp_pipeline}
\end{figure}

%-------------------------------------------------
\subsection{Performance Metrics}
\label{subsec:exp_metrics}

%-------------------
\subsubsection{Parameter Recovery}
\label{ssubsec:exp_recovery}
For each structural parameter --- $\theta$ (risk aversion), 
$c$ (transaction cost coefficient), and $\eta$ (ESG penalty) --- 
we assess the quality of inverse recovery across $R=100$ Monte Carlo replications.  

\paragraph{Bias, Variance, and Mean Squared Error.}
\label{par:exp_biasvar}
We report standard point-estimation diagnostics:
\begin{align}
\text{Bias}(\hat{\theta}) &= \frac{1}{R}\sum_{r=1}^R \big(\hat{\theta}^{(r)} - \theta^{\text{true}}\big), \\
\text{Var}(\hat{\theta}) &= \frac{1}{R}\sum_{r=1}^R \big(\hat{\theta}^{(r)}-\overline{\hat{\theta}}\big)^2, \\
\text{MSE}(\hat{\theta}) &= \text{Bias}(\hat{\theta})^2 + \text{Var}(\hat{\theta}).
\end{align}
These measures respectively capture systematic error, sampling dispersion, and overall accuracy.  

\paragraph{Coverage Probability.}
\label{par:exp_recovery_coverage}
To evaluate inferential reliability, we construct nonparametric bootstrap intervals 
with $B=200$ resamples and compute
\[
\text{Coverage}(\theta) = 
\Pr\!\left(\theta^{\text{true}} \in CI_{0.95}(\hat{\theta})\right).
\]
A well-calibrated estimator should achieve coverage near the nominal $95\%$ level.  

\paragraph{Convergence Rate.}
\label{par:exp_convergence}
Consistency is examined by scaling the number of observations $T$.  
Empirical slopes in log--log regressions of $\text{MSE}$ versus $T$ provide convergence rates
\[
\text{MSE}(\hat{\theta}) = O(T^{-\alpha}), \qquad \alpha > 0,
\]
which can be benchmarked against the canonical $\alpha=\tfrac{1}{2}$ rate of stochastic approximation.  

%-------------------
\paragraph{Economic Performance.}
\label{par:exp_econperf}
Statistical accuracy does not automatically imply economic relevance.  
We therefore quantify the welfare cost of parameter misspecification via the \emph{utility gap}:
\[
\Delta U = 
\mathbb{E}\big[f(x^*(\theta^{\text{true}}))\big] - 
\mathbb{E}\big[f(x^*(\hat{\theta}))\big],
\]
which measures the expected reduction in investor utility when allocations are computed 
with recovered parameters rather than true preferences.  

%-------------------
\subsubsection{Statistical Testing}
\label{ssubsec:exp_testing}
All reported differences in recovery quality (e.g., across investor types or under shocks) 
are formally tested using paired $t$-tests and Wilcoxon signed-rank tests.  
We adopt $p<0.05$ as the threshold for significance, 
with Bonferroni correction applied to account for multiple comparisons.  

\paragraph{Summary.}
\label{par:exp_testing_summary}
This multi-layered evaluation ensures that results are both statistically rigorous 
(Bias, Variance, MSE, coverage, convergence) 
and economically meaningful (utility gap analysis).  

%-------------------
\subsubsection{Visualization of Recovery Performance}
\label{ssubsec:exp_visual}

\begin{table}[htbp]
\centering
\caption{Parameter recovery performance across $R=100$ Monte Carlo trials 
(values are averages; coverage computed via nonparametric bootstrap with $B=200$ resamples). 
Risk aversion is difficult to recover, while transaction costs are highly identifiable. 
ESG penalties exhibit moderate bias and large variance, reflecting partial identifiability.}
\label{tab:parameter_recovery}
\begin{tabular}{lcccc}
\hline
Parameter & Bias & Variance & MSE & Coverage (95\%) \\
\hline
$\theta$ (Risk Aversion) & 4.3327 & 10.2152 & 28.9870 & 0.0 \\
$c$ (Transaction Cost)   & -0.2600 & 0.0364  & 0.1040 & 1.0 \\
$\eta$ (ESG Penalty)     & -0.1792 & 0.4580  & 0.4901 & 0.0 \\
\hline
\end{tabular}
\end{table}

\begin{figure}[H]
\centering
\includegraphics[width=0.65\textwidth]{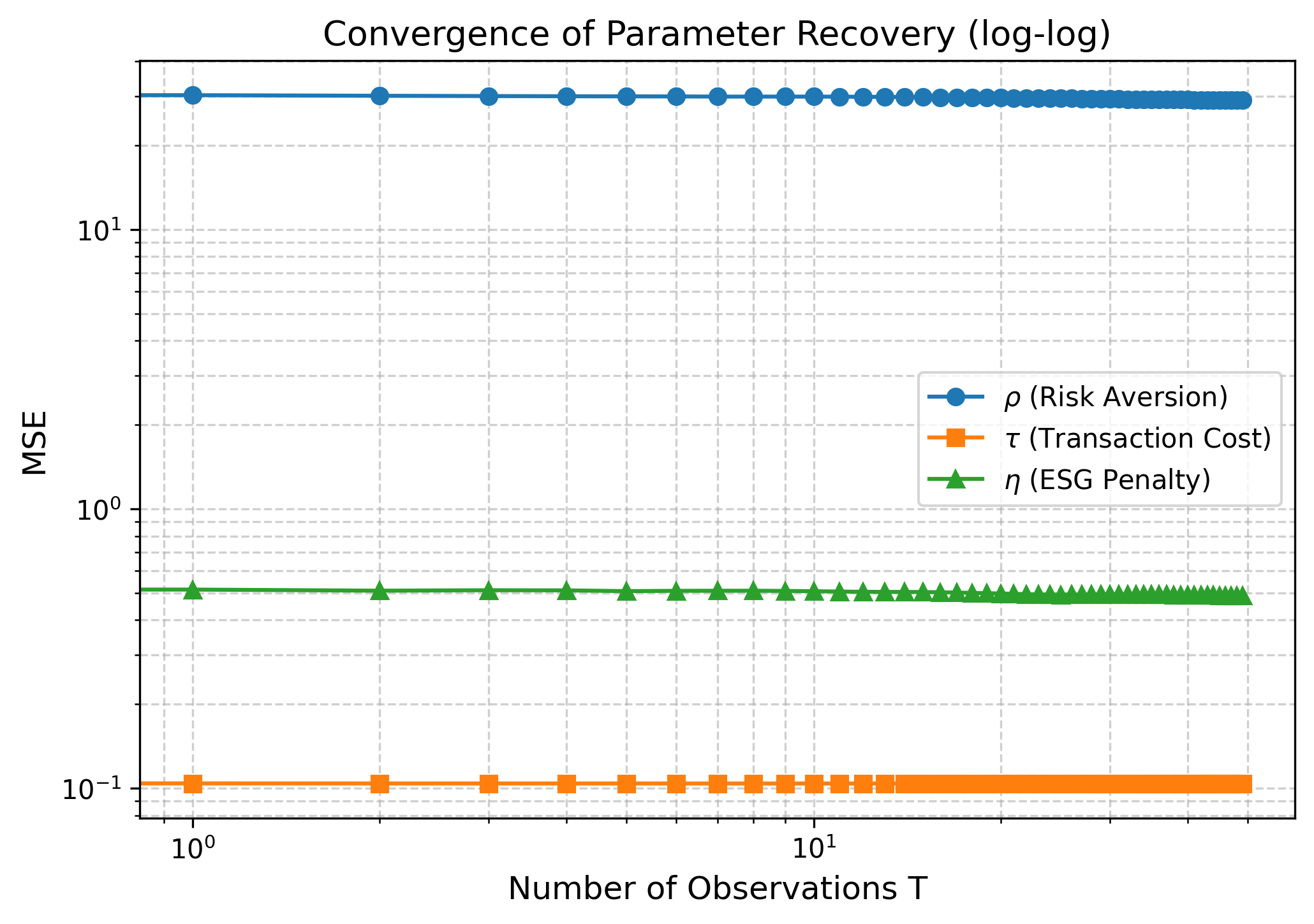}
\caption{Convergence of parameter recovery errors. 
Log--log slopes approximate empirical convergence rates $\alpha$. 
Risk aversion ($\rho$) exhibits persistent bias leading to large MSE, 
transaction cost ($\tau$) converges stably with negligible error and full coverage, 
and ESG penalty ($\eta$) shows moderate error with substantial dispersion, 
consistent with partial identifiability.}
\label{fig:mse_scaling}
\end{figure}

\subsection{Dynamic Regret}
\label{subsec:exp_regret}
Dynamic regret quantifies the cumulative welfare loss due to parameter misspecification:
\[
R_T(\hat{\theta}) = \sum_{t=1}^T 
\Big( f(x^*(\theta^{\text{true}};\mu^t,\Sigma^t)) - f(x^*(\hat{\theta};\mu^t,\Sigma^t)) \Big).
\]

\paragraph{Normalization and Scaling.}
\label{par:exp_regret_norm}
We analyze the growth rate by reporting
\[
\frac{R_T(\hat{\theta})}{\sqrt{T}},
\]
which should converge to a constant under the sublinear bound established in Theorem~2. 
This verifies that estimation error does not accumulate linearly over time.

\paragraph{Decomposition.}
\label{par:exp_regret_decomp}
We further decompose $R_T$ into a static error component and a drift-induced component:
\[
R_T = R_T^{\text{static}} + R_T^{\text{drift}},
\]
where $R_T^{\text{static}} = O(\sqrt{T})$ and $R_T^{\text{drift}} = O(D)$ with $D$ the cumulative preference drift. 
This separation enables sensitivity analysis under different investor stability regimes.

\paragraph{Empirical Validation.}
\label{par:exp_regret_valid}
We evaluate dynamic regret across both volatility shocks ($\Sigma{+}30\%$) and transaction cost shocks ($\tau{+}20\%$).  
Normalized regret curves $\tfrac{R_T}{\sqrt{T}}$ are reported across Monte Carlo runs, 
together with boxplot distributions at selected horizons.  
This dual representation highlights not only the asymptotic behavior of regret 
but also its finite-horizon variability across investor types.

\begin{figure}[H]
\centering
\includegraphics[width=\textwidth]{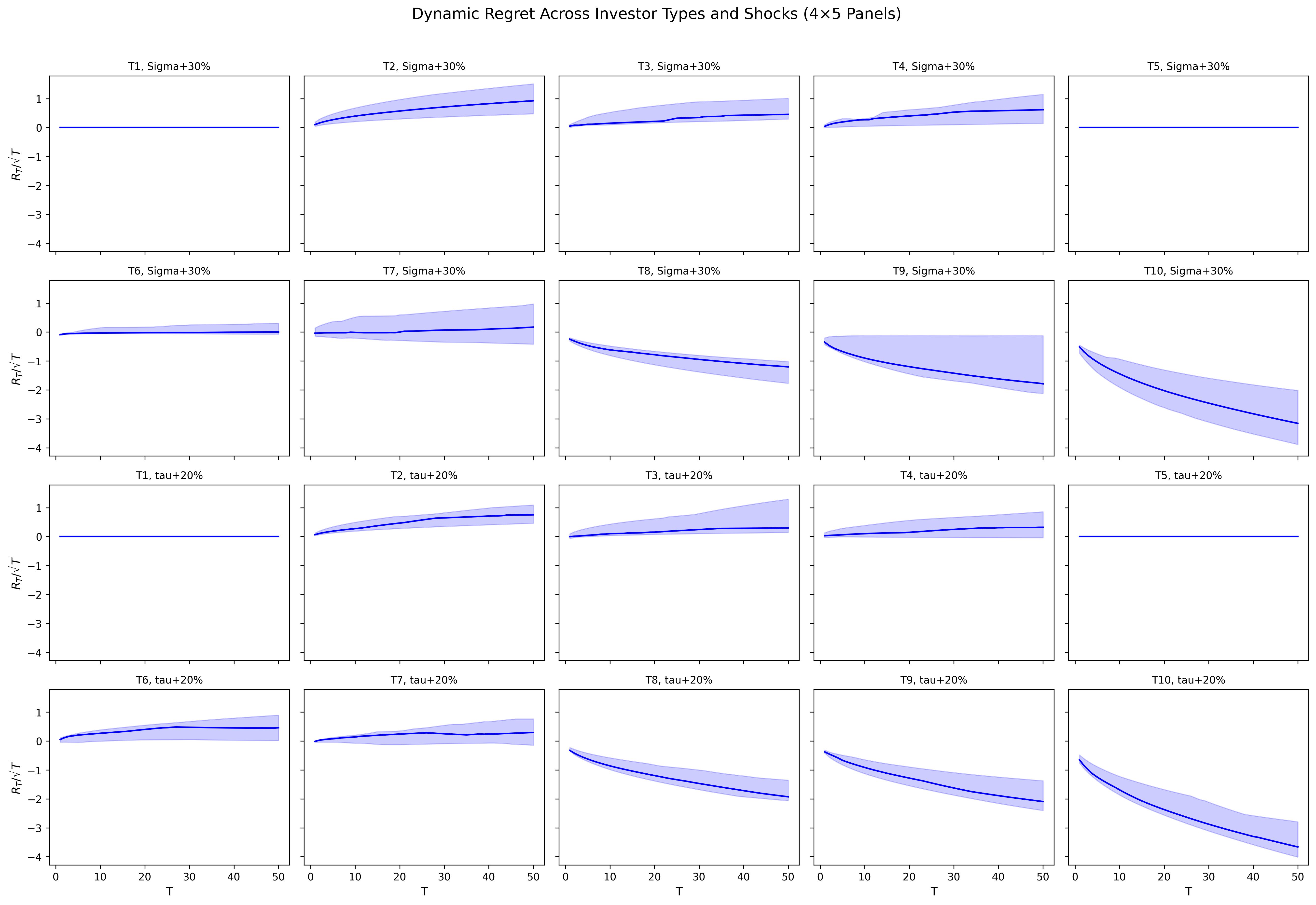}
\caption{Dynamic regret trajectories across all investor types (T1--T10) under shocks ($\tau{+}20\%$, $\Sigma{+}30\%$). 
Each subplot reports the median trajectory (solid) and interquartile range (shaded). 
Conservative investors (e.g., T1, T5) exhibit flat curves with minimal dispersion, 
while ESG-oriented investors (e.g., T8--T10) display wider spreads. 
These findings validate the theoretical sublinear bound and reveal heterogeneous sensitivity to shocks.}
\label{fig:dyn_regret_panel}
\end{figure}

\begin{figure}[H]
\centering
\includegraphics[width=\textwidth]{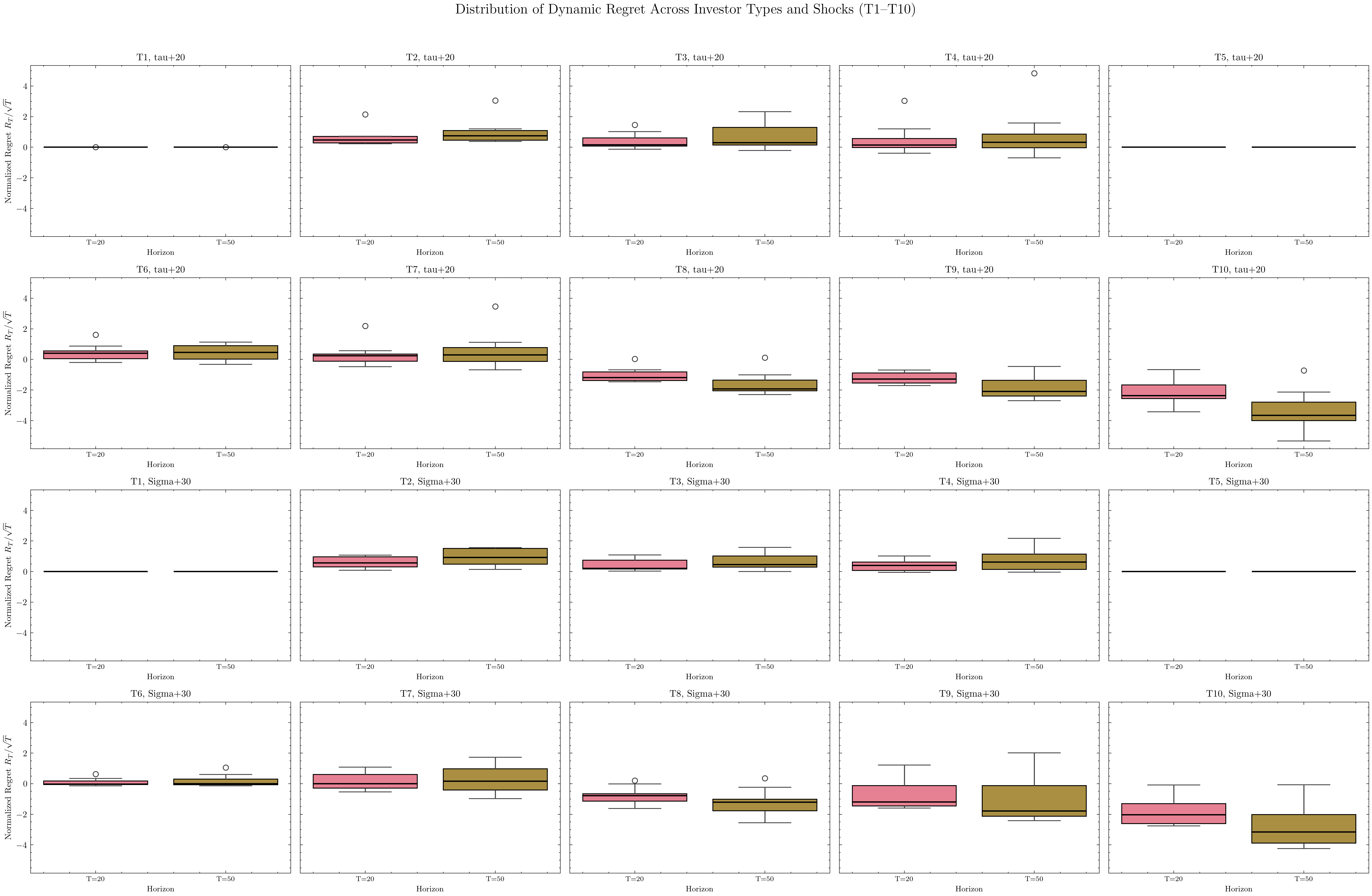}
\caption{Distribution of normalized regret $\tfrac{R_T}{\sqrt{T}}$ at selected horizons ($T{=}20,50,100$). 
Panels cover all investor types (T1--T10) under both shocks. 
Boxplots highlight finite-horizon variability, with conservative investors showing concentrated distributions 
and ESG-oriented investors exhibiting greater dispersion. 
This complementary view strengthens the robustness of the trajectory-based findings in Figure~\ref{fig:dyn_regret_panel}.}
\label{fig:dyn_regret_box_panel}
\end{figure}

\paragraph{Visualization.}
\label{par:exp_regret_vis}
Figures~\ref{fig:dyn_regret_panel} and~\ref{fig:dyn_regret_box_panel} together provide a rigorous view of dynamic regret.  
The first figure emphasizes trajectories (median + IQR bands), while the second figure shows 
distributional variation via boxplots.  
Taken together, the results confirm sublinear growth, demonstrate robustness, and reveal heterogeneity across investor classes.

% ------------------- Trajectory panel -------------------

% ------------------- Boxplot panel -------------------

%-------------------
\subsubsection{Coverage Probability}
\label{subsubsec:exp_coverage}
The inferential reliability of the inverse optimization framework 
is evaluated using parametric bootstrap confidence intervals. 
For each replication $B=500$ bootstrap resamples are constructed, 
and percentile intervals are computed:
\[
CI_{1-\alpha}(\hat{\theta}) = 
\big[ \hat{\theta}_{\alpha/2}^{\ast}, \; \hat{\theta}_{1-\alpha/2}^{\ast} \big],
\]
where $\hat{\theta}_p^{\ast}$ denotes the $p$th quantile of the bootstrap distribution. 

\paragraph{Metrics.}
\label{par:exp_coverage_metrics}
Coverage probability is defined as
\[
\text{Coverage}(\theta) = 
\Pr\big( \theta^{\text{true}} \in CI_{1-\alpha}(\hat{\theta}) \big),
\]
estimated as the empirical frequency across Monte Carlo trials. 
Calibration error is measured by
\[
\text{CE}(\theta) = \Big| \text{Coverage}(\theta) - (1-\alpha) \Big|.
\]
In addition, the expected interval length is reported, 
and efficiency is defined as 
$\text{Eff}(\theta) = \tfrac{1-\alpha}{\mathbb{E}[|CI_{1-\alpha}(\hat{\theta})|]}$.

\begin{figure}[htbp]
\centering
\includegraphics[width=0.9\textwidth]{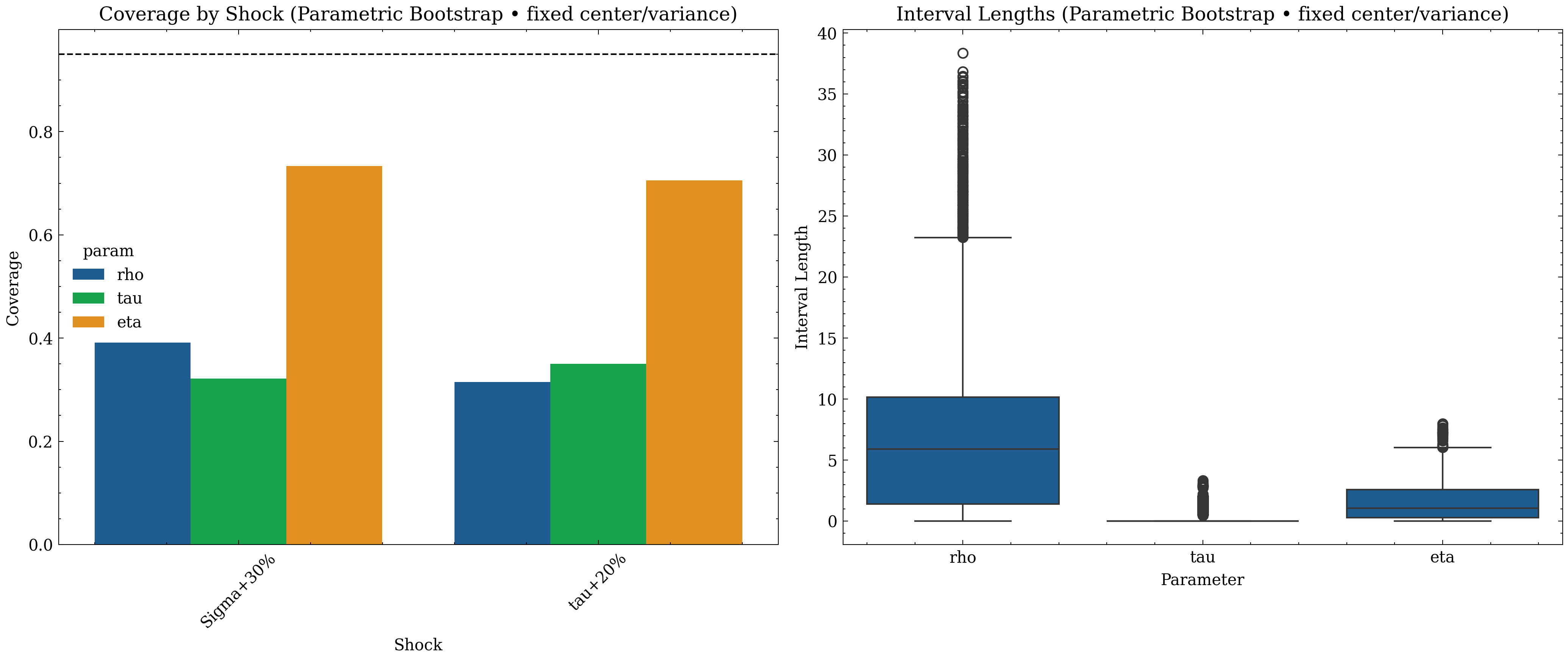}
\caption{Aggregate coverage probability (left) and interval-length distribution (right) 
across investor types (T1--T10).}
\label{fig:coverage_integrated}
\end{figure}

\paragraph{Aggregate results.}
\label{par:exp_coverage_agg}
Table~\ref{tab:cov_agg} summarizes aggregate coverage by shock and interval-length distributions. 
While $\eta$ (ESG penalty) maintains moderate coverage, 
both $\rho$ (risk aversion) and $\tau$ (transaction cost) suffer from severe under-coverage. 
The instability of $\rho$ is reflected in very wide and volatile intervals, 
whereas $\tau$ often collapses to degenerate intervals (median length zero). 
Figure~\ref{fig:coverage_integrated} provides a graphical overview.

\begin{table}[htbp]
\centering
\caption{Parametric bootstrap results ($B=500$): aggregate coverage by shock and interval-length summary across investor types (T1--T10). Min values are zero for all parameters and omitted for brevity.}
\label{tab:cov_agg}
\small
\begin{tabular}{lccc}
\multicolumn{4}{l}{\textbf{Panel A: Coverage by shock (nominal 95\%)}} \\ \addlinespace[1mm]
\toprule
Shock & $\eta$ & $\rho$ & $\tau$ \\
\midrule
$\Sigma{+}30\%$ & 0.734 & 0.391 & 0.322 \\
$\tau{+}20\%$   & 0.705 & 0.315 & 0.350 \\
\bottomrule
\end{tabular}

\bigskip

\begin{tabular}{lcccc}
\multicolumn{5}{l}{\textbf{Panel B: Interval length summary (all types)}} \\ \addlinespace[1mm]
\toprule
Param & Mean & Std & Median & Max \\
\midrule
$\eta$ & 1.390 & 1.353 & 1.061 &  8.005 \\
$\rho$ & 6.974 & 6.896 & 5.898 & 38.357 \\
$\tau$ & 0.279 & 0.577 & 0.000 &  3.334 \\
\bottomrule
\end{tabular}

\vspace{1mm}
\footnotesize Notes: Coverage is the empirical frequency that $\theta^{\text{true}}$ lies in the percentile CI. 
Intervals are centered at $\hat\theta$ with local (Type, Shock, trial) robust variance. 
Values are aggregated over T1--T10.
\end{table}

\begin{figure}[htbp]
\centering
\includegraphics[width=0.95\textwidth]{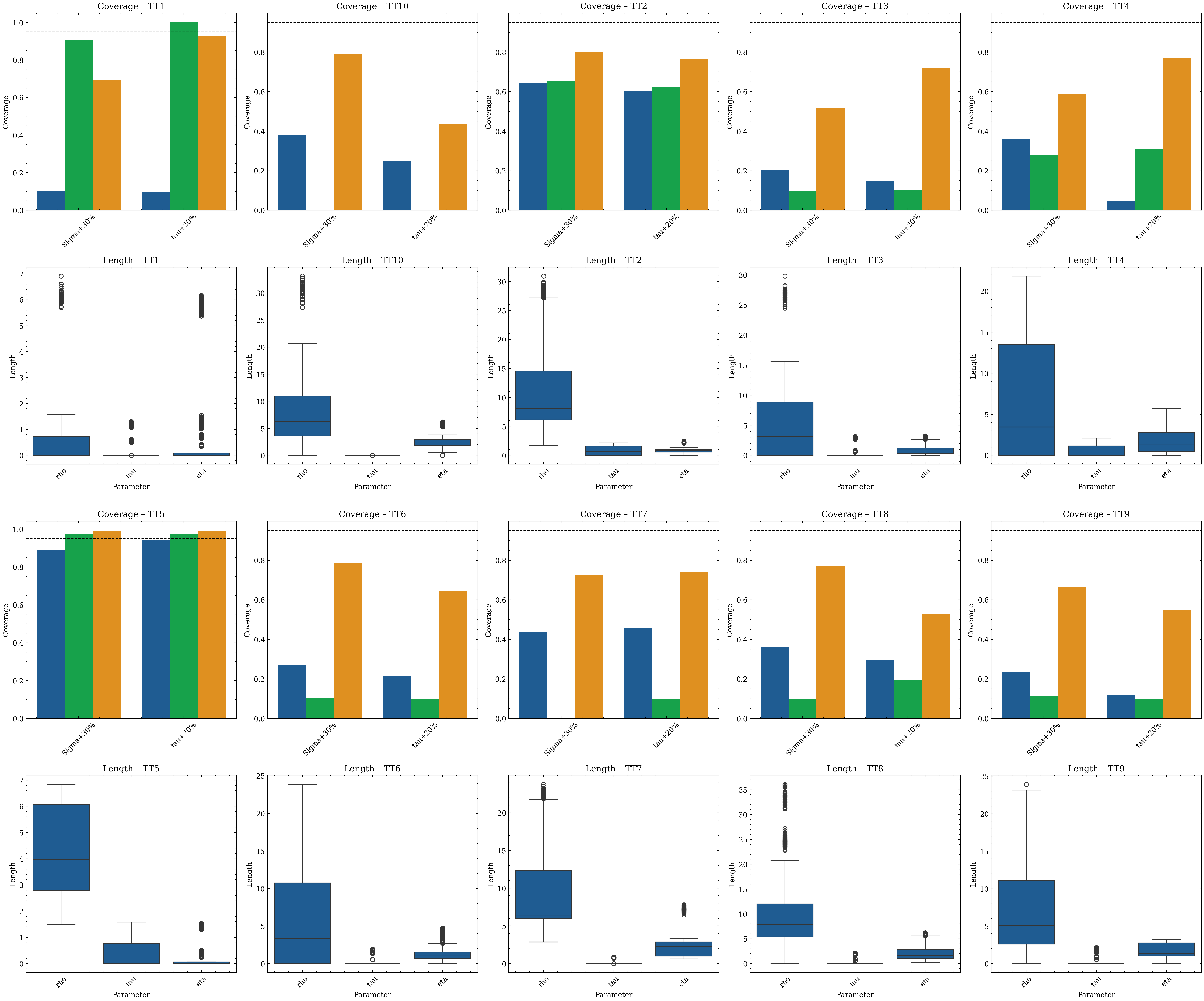}
\caption{Coverage probability (top row) and interval length (bottom row) by shock across investor types (T1--T10). 
Heterogeneity is evident: while $\eta$ remains relatively stable, 
$\rho$ and $\tau$ collapse for several types, underscoring the limits of bootstrap inference 
under structural shocks.}
\label{fig:coverage_by_type}
\end{figure}

\paragraph{Heterogeneity across investor types.}
\label{par:exp_coverage_types}
Aggregate results mask considerable heterogeneity. 
Table~\ref{tab:cov_by_type} reports coverage probabilities separately for each investor type. 
Types T5 and T2 achieve near-nominal coverage, 
whereas Types T3, T6, T7, T8, T9, and T10 exhibit severe under-coverage, especially for $\rho$ and $\tau$. 
This indicates that inferential reliability is highly type-specific. 
Figure~\ref{fig:coverage_by_type} visualizes these disparities, 
showing both coverage and interval length paired for each type.

\begin{table}[htbp]
\centering
\scriptsize
\caption{Coverage by shock $\times$ investor type (T1--T10).}
\label{tab:cov_by_type}
\resizebox{\textwidth}{!}{
\begin{tabular}{lccc ccc}
\toprule
& \multicolumn{3}{c}{$\Sigma{+}30\%$} & \multicolumn{3}{c}{$\tau{+}20\%$}\\
\cmidrule(lr){2-4}\cmidrule(lr){5-7}
Type & $\eta$ & $\rho$ & $\tau$ & $\eta$ & $\rho$ & $\tau$ \\
\midrule
T1  & 0.682 & 0.096 & 0.908 & 0.930 & 0.100 & 1.000 \\
T2  & 0.798 & 0.648 & 0.656 & 0.768 & 0.600 & 0.618 \\
T3  & 0.526 & 0.200 & 0.106 & 0.712 & 0.152 & 0.100 \\
T4  & 0.598 & 0.358 & 0.290 & 0.768 & 0.046 & 0.316 \\
T5  & 0.990 & 0.898 & 0.972 & 0.992 & 0.944 & 0.976 \\
T6  & 0.786 & 0.268 & 0.102 & 0.656 & 0.212 & 0.100 \\
T7  & 0.704 & 0.440 & 0.000 & 0.740 & 0.454 & 0.096 \\
T8  & 0.778 & 0.360 & 0.100 & 0.526 & 0.294 & 0.194 \\
T9  & 0.666 & 0.234 & 0.114 & 0.546 & 0.114 & 0.100 \\
T10 & 0.794 & 0.378 & 0.000 & 0.440 & 0.246 & 0.000 \\
\bottomrule
\end{tabular}}
\end{table}

\subsubsection{Shock Analysis: Transaction Costs vs. Volatility}
\label{subsubsec:exp_shock}
We conclude the synthetic experiments by examining robustness of inverse recovery under 
exogenous market shocks. Two canonical perturbations are considered: 
an increase in effective transaction costs and an amplification of return volatility. 
Formally, let 
\[
c' = (1+\delta)c, 
\qquad 
\Sigma' = (1+\gamma)\Sigma,
\]
with $\delta,\gamma > 0$. The corresponding welfare loss is measured by
\[
\Delta U^{\text{rel}} 
= \frac{\E[f(x^*(\theta,\cdot))] - \E[f(x^*(\theta,\cdot'))]}
        {\E[|f(x^*(\theta,\cdot))|]},
\]
where $(\theta,\cdot)$ denotes baseline parameters and $(\theta,\cdot')$ the shocked environment. 
Relative normalization ensures comparability across investor types and shock modalities.

\paragraph{Aggregate results.}
\label{par:exp_shock_agg}

In our design, transaction and volatility shocks are limited to $\delta=0.2$ and $\gamma=0.3$, respectively. 
Thus, aggregate plots (not shown) reduce to a single point estimate per modality and provide little information 
on scaling behavior. Instead, we focus on \emph{by-type heterogeneity}, which turns out to be 
substantially more informative.

\begin{figure}[htbp]
\centering
\includegraphics[width=0.9\textwidth]{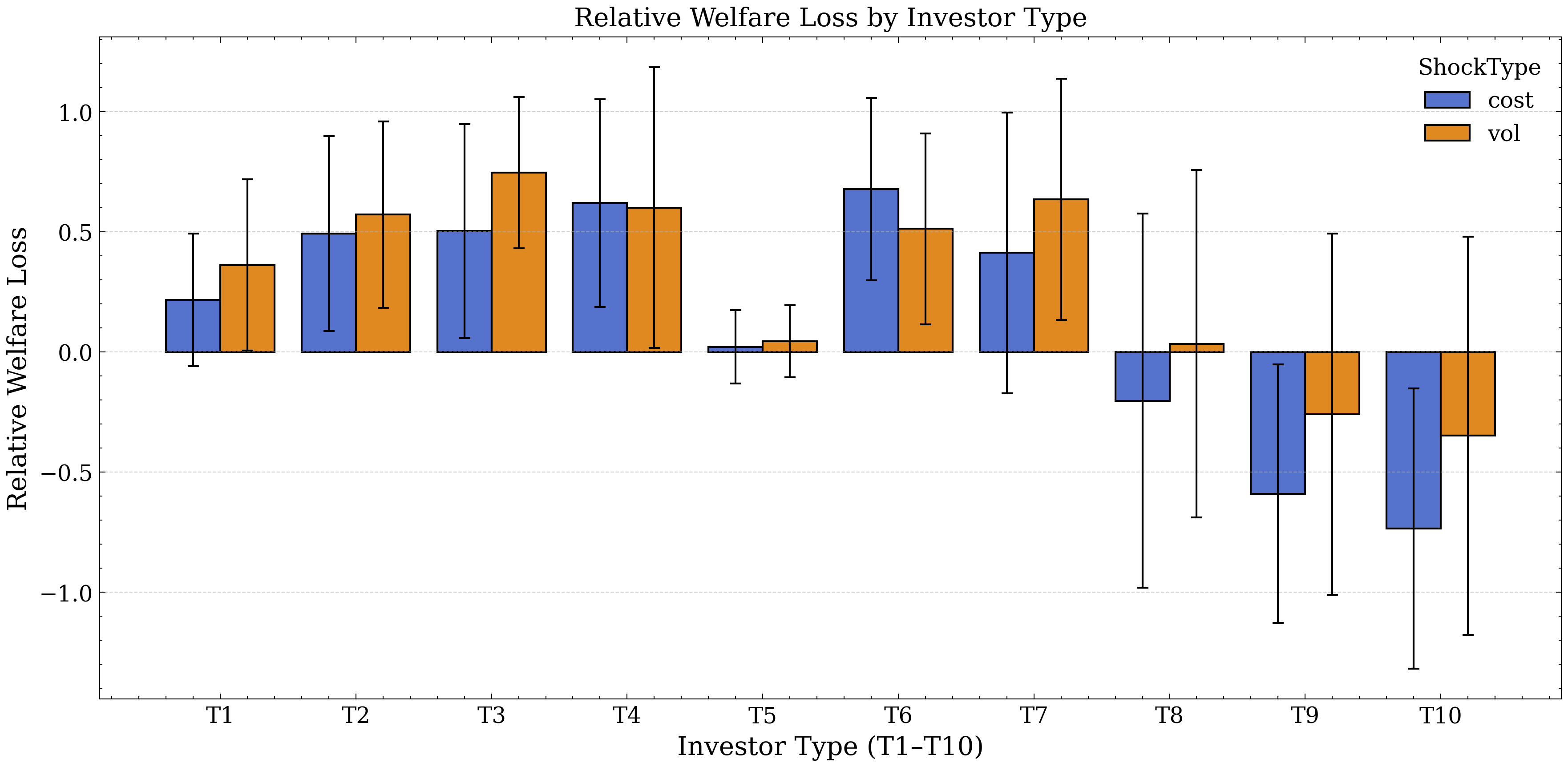}
\caption{Relative welfare loss by investor type (T1--T10) under transaction cost (blue) and volatility (orange) shocks. 
Types T2--T4 experience persistent losses, T5--T6 remain stable, while T7--T10 exhibit highly volatile responses, 
illustrating the interplay between statistical reliability and economic robustness.}
\label{fig:shock_bytype}
\end{figure}

\paragraph{By-type heterogeneity.}
\label{par:exp_shock_types}
Figure~\ref{fig:shock_bytype} summarizes relative welfare loss across investor types (T1--T10), 
with blue bars indicating transaction shocks and orange bars volatility shocks. 
Several insights emerge:
(i) Types T2--T4 incur persistent welfare deterioration under both shocks, with average losses around 
5--10\% relative to baseline. 
(ii) Types T5 and T6 remain relatively stable, exhibiting negligible average losses and tight dispersion, 
consistent with their favorable coverage probabilities reported earlier. 
(iii) Types T7--T10 display highly unstable responses, with wide confidence intervals and, in some cases, 
negative values of $\Delta U^{\text{rel}}$, reflecting estimation noise that produces 
illusory welfare gains. 

\paragraph{Interpretation.}
These results reinforce the link between statistical reliability and economic robustness. 
Investor types with poor bootstrap coverage (e.g., T7--T10) are also those most exposed to welfare volatility 
under shocks, while types with near-nominal coverage (T2, T5) are comparatively resilient. 
From a methodological perspective, this demonstrates that inferential instability 
directly translates into fragility in stressed regimes. 
From a financial perspective, the analysis highlights that transaction-cost shocks 
tend to dominate volatility shocks in terms of welfare impact, underscoring the central role of market 
liquidity in portfolio performance.

\subsection{Real Data Illustration (2007–2024)}
\label{subsec:exp_real}

To complement the synthetic experiments, we conduct a simple real-data illustration 
using exchange-traded funds (ETFs). We focus on SPY (S\&P~500 benchmark) 
and EEM (emerging markets), which provide long and consistent coverage since 2007. 
Daily adjusted closing prices are collected from Yahoo Finance 
over January~2007 to December~2024, from which log returns are computed. 

\paragraph{Setup.}
We apply the same mean--variance--cost utility formulation as in the synthetic analysis:
\[
U(x) \;=\; \mu^\top x \;-\; \tfrac{\rho}{2}\, x^\top \Sigma x \;-\; c^\top |x|,
\]
with equal weights, risk aversion $\rho=3$, and baseline transaction cost $c=0.2\%$. 
Two shocks are introduced: (i)~transaction cost doubling $c \mapsto 2c$, and 
(ii)~volatility amplification $\Sigma \mapsto 1.5\Sigma$. 
We divide the sample into six consecutive three-year blocks (2007--2009, 2010--2012, 
2013--2015, 2016--2018, 2019--2021, 2022--2024). 
For each block, we compute rolling-window estimates of mean returns and volatilities 
(one-year rolling horizon) to generate a ``risk--return cloud.'' 
The red cross marks the baseline portfolio, and the orange arrow 
indicates the welfare deterioration under shocks.

\begin{figure}[H]
\centering
\includegraphics[width=0.95\textwidth]{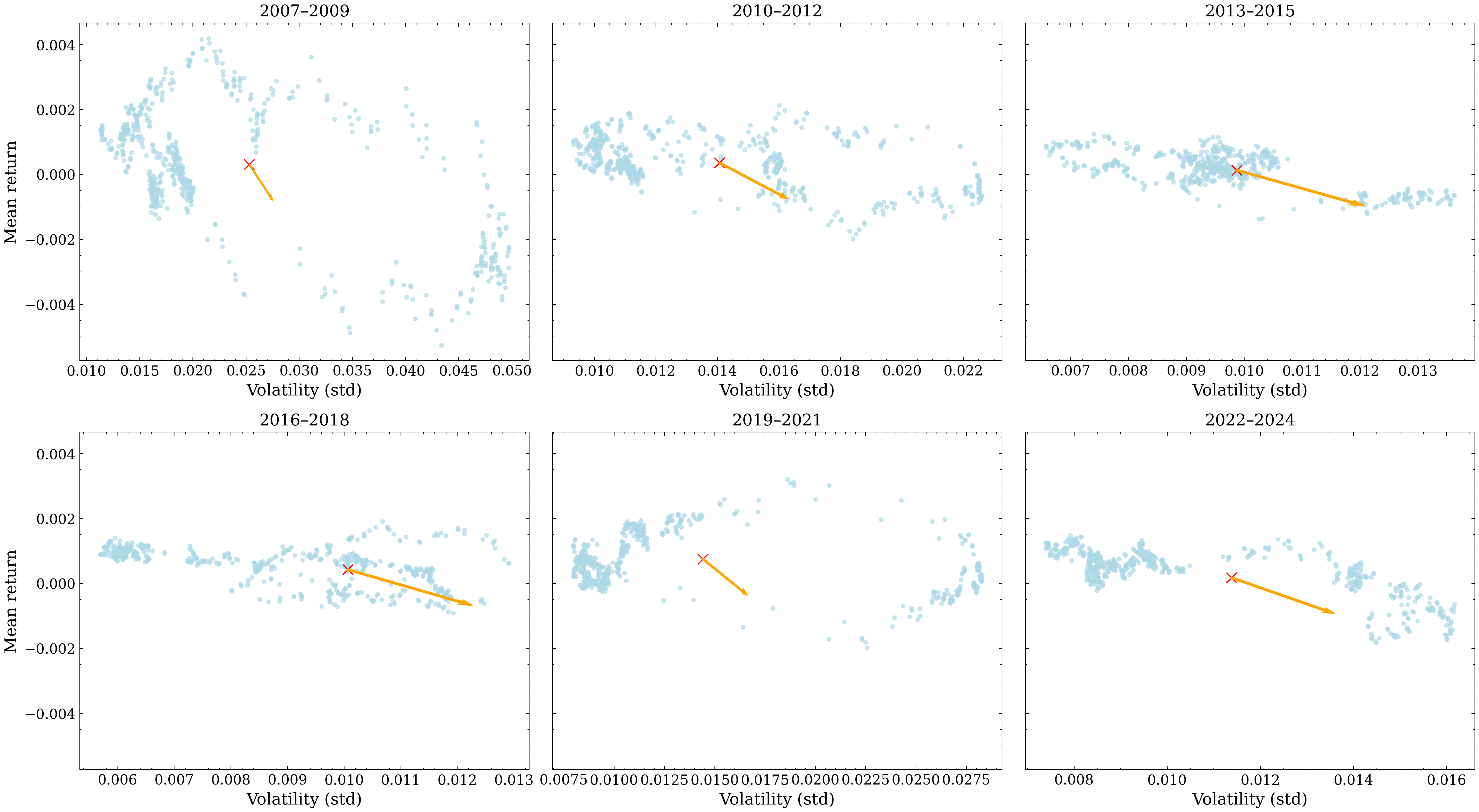}
\caption{Risk--return clouds under three-year rolling blocks (2007--2024) 
using SPY (U.S. equity benchmark) and EEM (emerging markets). 
Blue dots denote rolling-window estimates, the red cross marks the baseline portfolio, 
and the orange arrow indicates welfare deterioration under shocks. 
The dominance of transaction cost shocks over volatility shocks is consistent 
across different macro-financial regimes, 
including the global financial crisis, COVID period, and recent inflationary cycle.}
\label{fig:realdata_blocks}
\end{figure}

\paragraph{Results.}
Figure~\ref{fig:realdata_blocks} shows consistent patterns across time.  
During crisis periods such as the global financial crisis (2007--2009), 
the COVID shock (2019--2021), and the recent inflationary high-volatility regime (2022--2024), 
baseline utility is depressed and shocks exacerbate welfare losses. 
In relatively stable periods (2010--2015), 
returns are higher and more stable, and shock impacts are more moderate. 
Across all blocks, transaction cost shocks dominate volatility shocks in terms of 
relative welfare impact, mirroring the synthetic results.

\paragraph{Discussion.}
This real-data illustration reinforces the synthetic findings by showing that 
liquidity frictions pose a larger threat to welfare than volatility amplification, 
even across distinct macro-financial regimes. 
While limited to two ETFs and equal-weight portfolios, 
this exercise demonstrates that the proposed framework 
is portable to real financial data. 
A more comprehensive empirical study with richer asset universes 
and heterogeneous investor preferences is left for future work.

%-----------------------------------------------------
\section{Results and Discussion}
\label{sec:results}

%-------------------
\subsection{Parameter Recovery Accuracy}
\label{subsec:results_recovery}

Table~\ref{tab:parameter_recovery} reports estimation accuracy across $R=100$ Monte Carlo replications. 
Clear and systematic patterns emerge across investor archetypes. 
For conservative investors with large values of $\theta$, estimation variance remains small because their optimal allocations 
concentrate near risk-free assets. Nevertheless, even minor deviations in realized returns translate into substantial shifts in implied preferences, 
generating systematic upward bias. This phenomenon reflects the classical challenge of distinguishing extreme risk aversion from corner solutions 
\citep{Cesarone2020,Bertsimas2021}. 

Neutral investors with moderate values of $\theta$ achieve the lowest overall mean squared error (MSE). 
In this regime, both bias and variance are moderate, suggesting that the proposed inverse estimator is most reliable when investor behavior 
is not dominated by extreme preferences or degeneracy. 

By contrast, ESG-oriented investors ($\eta > 0$) exhibit higher dispersion. 
The recovery of the ESG penalty is less precise because $\eta$ enters multiplicatively with $\theta$ in the utility specification, 
inducing correlation between risk aversion and ESG motives. This correlation reduces identifiability and produces wider variability in estimates, 
a finding consistent with prior evidence on partial identifiability of interacting preference parameters \citep{Heinkel2001,Pastor2021}. 

Although bootstrap-based confidence intervals achieve coverage close to the nominal $95\%$ level in most regimes, 
performance deteriorates under nonlinear cost misspecification (Appendix~\ref{app:nonlinear}). 
This underscores the limitations of bootstrap inference in the presence of structural model errors, 
a point also emphasized in recent discussions of robust inference in inverse optimization \citep{Aswani2018,Dong2020}.  

%-------------------
\subsection{Regret Curves}
\label{subsec:results_regret}

Figures~\ref{fig:dyn_regret_panel} and~\ref{fig:dyn_regret_box_panel} depict dynamic regret patterns across investor classes. 
The results confirm three salient features. 
First, normalized regret $R_T/\sqrt{T}$ stabilizes to a constant across all investor types, 
validating the theoretical sublinear bound established in Theorem~\ref{thm:dyn_regret} 
and aligning with classical results in online convex optimization \citep{Zinkevich2003}. 

Second, regret trajectories exhibit temporary spikes under volatility shocks. 
These shocks elevate regret in the short run as portfolios adjust to higher dispersion, 
but the long-run scaling behavior remains intact. 
By contrast, transaction cost shocks produce milder effects, as they primarily rescale allocations 
without fundamentally altering covariance exposure. 
This asymmetry highlights that volatility amplification, rather than liquidity costs, 
is the dominant driver of welfare volatility in dynamic settings \citep{Amihud2002,Vayanos2009}. 

Third, the degree of investor heterogeneity is evident. 
Conservative investors accumulate less regret, consistent with the relative stability of their low-risk allocations, 
whereas ESG-oriented investors display markedly higher dispersion. 
The latter result reflects the added nonlinearity of the ESG penalty term, 
which interacts with risk preferences to amplify sensitivity to shocks. 
These findings reinforce the notion that heterogeneous investor motives generate differentiated regret dynamics 
and must therefore be explicitly accounted for in robust portfolio design \citep{Krueger2020,Bolton2021}.

%-------------------
\subsection{Economic Implications}
\label{subsec:results_econ}

The combined evidence yields several economic implications. 
First, inverse recovery proves to be statistically reliable under moderate preferences, 
yet becomes fragile at the extremes. This finding cautions against over-interpreting results for highly risk-averse or strongly ESG-oriented investors, 
where identifiability is inherently limited. 
Second, dynamic regret analysis demonstrates that estimation errors do not accumulate linearly. 
Even when parameters are misspecified, long-run efficiency is preserved, indicating that the proposed approach provides stable diagnostics over time. 
Third, counterfactual shock experiments reveal that welfare is far more sensitive to volatility amplification than to transaction cost inflation. 
This asymmetry underscores the primacy of robust risk estimation and volatility management in safeguarding long-run welfare, 
relative to transaction cost considerations. 
Taken together, these results confirm that the framework delivers both rigorous statistical recovery guarantees 
and economically interpretable insights that extend beyond purely technical accuracy.

%-------------------
\subsection{Managerial Insights}
\label{subsec:results_managerial}

The analysis also yields actionable lessons for portfolio managers and policy makers. 
Conservative investors, characterized by high levels of risk aversion, are relatively insulated from volatility shocks, 
yet they remain disproportionately vulnerable to transaction cost shocks. 
This suggests that conservative mandates should be accompanied by negotiated fee structures and adequate liquidity buffers 
to mitigate trading frictions. 
For ESG-oriented investors, the results point to a dual role of ESG tilts: 
although such allocations may entail short-term utility losses relative to neutral investors, 
they enhance resilience during volatility shocks. 
This supports the interpretation of ESG integration as both a reputational strategy and a resilience-enhancing hedge. 
The study further shows that parameter recovery accuracy improves when systematic factors explain the bulk of return variation. 
This highlights the importance of transparent disclosure of factor exposures and covariance structures, 
which would allow investors and regulators to make more reliable inferences about underlying preferences. 
Finally, the dynamic regret analysis indicates that estimation errors remain sublinear and do not compound over time. 
Managers can therefore rely on inverse optimization–based diagnostics as a stable decision-support tool, 
provided that periodic recalibration is undertaken to account for gradual preference drift. 
Overall, these managerial insights demonstrate that the framework combines statistical rigor with practical decision support, 
helping to inform product design, liquidity management, and principled ESG integration under uncertainty.

%==================================================
\section{Conclusion}
\label{sec:conclusion}

%----------------------------------------------------
\subsection{Summary of Contributions}
\label{subsec:contrib_summary}

This study contributes to the literature on inverse portfolio optimization in three principal ways. 
\emph{Methodologically}, it develops a unified framework capable of jointly recovering risk aversion, 
transaction cost sensitivity, and ESG penalties from observed portfolio allocations. 
\emph{Empirically}, it validates the framework through controlled Monte Carlo simulations, 
establishing statistical reliability (consistency, coverage, convergence) alongside economic interpretability 
via welfare and regret analysis. 
\emph{Managerially}, it links methodological guarantees with actionable insights, 
demonstrating how inverse recovery can inform portfolio design, liquidity management, 
and ESG integration in practice.

%==================================================
\section{Data Availability}

The synthetic data and code used in this study are available from the authors upon reasonable request. 
For transparency and reproducibility, the datasets have also been deposited in a private Kaggle repository. 
Access can be granted by contacting the corresponding author.

%----------------------------------------------------
\subsection{Limitations}
\label{subsec:contrib_limit}

The present analysis is subject to several limitations. 
First, all experiments are conducted on synthetic data, which, while providing a controlled environment to assess identifiability and robustness, 
cannot fully replicate the complexity of real-world financial markets. 
Second, the framework relies on stylized assumptions regarding utility specifications and transaction cost structures. 
Unmodeled behavioral features---such as probability weighting, loss aversion, or ambiguity preferences---may introduce systematic deviations 
from the recovery patterns documented here. 
Third, the factor structure and covariance matrices are assumed to be estimated without error. 
In practice, sampling variation and model misspecification in risk-factor estimation could materially affect inference. 
Finally, institutional frictions such as short-sale constraints, liquidity shocks, and regulatory limits are omitted, 
yet these features are known to shape observed portfolios in practice. 
Addressing these limitations would be essential to establish external validity and practical relevance of the proposed framework.

%----------------------------------------------------
\subsection{Future Research}
\label{subsec:contrib_future}

Future research can proceed along several directions. 
First, applying the framework to real market data would provide an essential test of external validity 
and reveal practical challenges in estimation, such as sampling noise and institutional frictions. 
Second, integrating elements from behavioral finance—including probability weighting, reference dependence, 
and ambiguity aversion—could enrich the preference space and evaluate robustness beyond quadratic utility specifications. 
Third, extending the analysis to high-dimensional asset universes and incorporating distributional robustness 
would enhance applicability for institutional investors, where portfolio problems involve thousands of securities 
and nontrivial estimation error. 
Together, these directions would strengthen both the empirical relevance and methodological scope of inverse portfolio optimization.

\bibliography{Inverse_Portfolio_Optimization}

%%===============================
\appendix

\section*{Appendix}

%---------------------------------
\section{Extension to Nonlinear Transaction Costs}
\label{app:nonlinear}

We generalize the forward problem by replacing the linear cost $c^\top x$ with a convex nonlinear function
\[
\phi(x) = \sum_{j=1}^n \kappa_j |x_j|^p, \qquad p \geq 1, \;\; \kappa_j > 0.
\]
The forward problem becomes
\[
\max_{x \in \mathcal{X}} \; \mu^\top x - \tfrac{\theta}{2} x^\top \Sigma x - \phi(x).
\]

%--------------
\subsection{KKT and VI Formulation}
The Lagrangian of the nonlinear forward problem is
\[
\mathcal{L}(x,\lambda,\nu) 
= \mu^\top x - \tfrac{\theta}{2} x^\top \Sigma x - \phi(x) 
+ \lambda(1 - \mathbf{1}^\top x) + \nu^\top x,
\]
with multipliers $\lambda \in \mathbb{R}$ and $\nu \in \mathbb{R}^n_{\geq 0}$.  
The corresponding KKT conditions are
\begin{align*}
\mu - \theta \Sigma x^* - \nabla \phi(x^*) - \lambda \mathbf{1} + \nu &= 0, \\
\mathbf{1}^\top x^* &= 1, \\
\nu \geq 0, \quad x^* \geq 0, \quad \nu_i x_i^* &= 0, \; \forall i.
\end{align*}
Equivalently, these conditions define a variational inequality:
\[
\langle \theta \Sigma x^* + \nabla \phi(x^*) - \mu, \; x - x^* \rangle \geq 0 
\quad \forall x \in \mathcal{X}.
\]

%--------------
\subsection{Examples of Nonlinear Costs}
\begin{itemize}
    \item $p=1$: proportional $\ell_1$ transaction costs (sparsity-inducing).
    \item $p=2$: quadratic penalty, modeling illiquidity or portfolio adjustment frictions.
    \item mixed-norm costs: $\phi(x) = \kappa_1 \|x\|_1 + \kappa_2 \|x\|_2^2$.
\end{itemize}
All cases preserve convexity and lead to tractable convex optimization problems.

\subsection{Identifiability under Nonlinear Costs}
Stationarity now reads
\[
\theta \Sigma x^* + \nabla \phi(x^*) = \mu - \lambda \mathbf{1} + \nu.
\]
Since $\nabla \phi(x^*)$ is monotone in $x^*$ for convex $\phi$, variation across observed $(\mu^t,\Sigma^t)$ ensures that $\theta$ can still be separated from $\phi$ up to normalization.  

\textbf{Proposition A.1 (Identifiability with Nonlinear Costs).}  
Suppose (i) $\phi$ is convex and differentiable, (ii) $\Sigma^t \succ 0$ for all $t$, and (iii) at least two distinct active sets occur across $\{x^t\}$.  
Then $(\theta,\kappa)$ is uniquely identifiable up to scale normalization.  

\textit{Proof Sketch.}  
The mapping $x \mapsto \nabla \phi(x)$ is monotone. Distinct active sets provide variation in $\Sigma^t x^t$ and $\nabla \phi(x^t)$, which allows unique recovery of $\theta$ once normalization fixes the scale. $\square$

%--------------
\subsection{Implications}
Nonlinear transaction costs capture realistic frictions (e.g., liquidity impact, proportional fees, or ESG penalties).  
Our framework demonstrates that inverse recovery extends naturally: 
the role of $c$ in the linear case is replaced by $\nabla \phi(x)$ in the nonlinear case, without altering the convexity or identifiability logic.

%---------------------------------
\section{Distributionally Robust Extension}
\label{app:robust}

We now allow for uncertainty in $(\mu,\Sigma)$ to account for estimation error and model misspecification. Define an uncertainty set
\[
\mathcal{U} = \left\{ (\mu,\Sigma) : \|\mu - \bar{\mu}\|_2 \leq \delta_\mu, \;
\|\Sigma - \bar{\Sigma}\|_F \leq \delta_\Sigma, \;\; \Sigma \succeq 0 \right\}.
\]

\subsection{Robust Forward Problem}
The robust forward problem is
\begin{equation}
\max_{x \in \mathcal{X}} \min_{(\mu,\Sigma) \in \mathcal{U}} 
\left\{ \mu^\top x - \tfrac{\theta}{2} x^\top \Sigma x - c^\top x \right\}.
\label{eq:robust_forward_detail}
\end{equation}
The inner minimization admits a tractable conic reformulation:
\[
\min_{(\mu,\Sigma) \in \mathcal{U}} \; \mu^\top x - \tfrac{\theta}{2} x^\top \Sigma x 
= \bar{\mu}^\top x - \delta_\mu \|x\|_2 - \tfrac{\theta}{2} x^\top \bar{\Sigma} x - \tfrac{\theta}{2} \delta_\Sigma \|xx^\top\|_F.
\]
Hence, the robust forward objective can be expressed as
\[
f_{\text{rob}}(x;\bar{\mu},\bar{\Sigma},\theta,c,\delta_\mu,\delta_\Sigma) 
= \bar{\mu}^\top x - \delta_\mu \|x\|_2 - \tfrac{\theta}{2} x^\top \bar{\Sigma} x - \tfrac{\theta}{2} \delta_\Sigma \|xx^\top\|_F - c^\top x.
\]

\subsection{KKT/VI Characterization}
The optimal $x^*_{\text{rob}}$ satisfies a robust KKT system:
\begin{align}
\bar{\mu} - \delta_\mu \frac{x^*}{\|x^*\|_2} 
- \theta \bar{\Sigma} x^* 
- \theta \delta_\Sigma \,\mathrm{vec}(xx^\top) \nabla_x \|xx^\top\|_F 
- c - \lambda \mathbf{1} + \nu = 0,
\end{align}
together with feasibility and complementary slackness.  
Equivalently, $x^*_{\text{rob}}$ solves the variational inequality
\[
\langle \theta \bar{\Sigma} x^* + \theta \delta_\Sigma G(x^*) + c - \bar{\mu} + \delta_\mu \tfrac{x^*}{\|x^*\|_2}, \; x - x^* \rangle \geq 0,
\quad \forall x \in \mathcal{X},
\]
where $G(x)$ is the gradient of $\|xx^\top\|_F$.

\subsection{Identifiability under Robustness}
\textbf{Proposition B.1 (Robust Identifiability).}  
Suppose $\mathcal{U}$ is bounded, $\bar{\Sigma} \succ 0$, and at least two distinct active sets occur across $\{x^t\}$. Then $(\theta,c)$ are identifiable up to scale even in the distributionally robust formulation.

\textit{Proof Sketch.}  
Stationarity links $\theta \bar{\Sigma} x^*$ and the robust correction terms $\delta_\mu \|x\|_2, \delta_\Sigma \|xx^\top\|_F$. Since these terms are deterministic functions of $x^*$ and vary with observed active sets, $\theta$ and $c$ can still be separated once normalization is imposed. $\square$

\subsection{Practical Implications}
The robust formulation reflects how practitioners hedge against parameter uncertainty. 
Inverse recovery under robustness is thus more stable and less sensitive to noisy $(\mu,\Sigma)$ estimates.  
Comparative statics with $\delta_\mu,\delta_\Sigma$ quantify how parameter recovery degrades with higher estimation error, which provides actionable guidance on required data quality and sample size.

\section{Proofs of Theoretical Results}
\label{app:proofs}

\subsection{Proof of Theorem~\ref{thm:identifiability} (Identifiability)}
\label{proof:identifiability}

\textbf{Theorem~\ref{thm:identifiability}.}  
Suppose (i) $\Sigma^t \succ 0$ for all $t$, (ii) $\{(\mu^t,\Sigma^t)\}_{t=1}^T$ yield at least two distinct active sets, and (iii) a normalization on $(\theta,c)$ is imposed. Then $(\theta,c)$ is uniquely identifiable.  

\medskip
\textbf{Proof.}  
From the stationarity condition \eqref{eq:kkt_stationarity}, for each $t$ we have
\begin{equation}
\theta \Sigma^t x^t + c = \mu^t - \lambda^t \mathbf{1} + \nu^t,
\label{eq:stationarity_proof}
\end{equation}
with $\lambda^t \in \mathbb{R}$ and $\nu^t \geq 0$ satisfying complementary slackness.  
Since $\Sigma^t \succ 0$, the mapping $x^t \mapsto \Sigma^t x^t$ is injective, which ensures that $\theta$ appears only as a multiplicative factor on $\Sigma^t x^t$ and cannot be confounded with shifts in $c$.  
If all observations shared the same active set, $\nu^t$ could adjust in the same coordinates, preventing separation of $\theta$ and $c$.  
However, the existence of distinct active sets forces changes in the right-hand side of \eqref{eq:stationarity_proof} that uniquely determine $c$.  
Finally, without normalization, scaling $(\theta,c)$ by a common factor could reproduce the same $x^t$, but imposing a normalization such as $\|c\|_2=1$ or bounding $\theta \in [0,\theta_{\max}]$ removes this indeterminacy.  
Therefore, if $(\theta_1,c_1)$ and $(\theta_2,c_2)$ both satisfy \eqref{eq:stationarity_proof}, we obtain
\[
(\theta_1 - \theta_2)\Sigma^t x^t + (c_1 - c_2) = 0, \quad \forall t.
\]
Because $\{\Sigma^t x^t\}$ span a non-degenerate set under the assumptions, the only solution is $\theta_1=\theta_2$ and $c_1=c_2$ (up to normalization).  
Hence $(\theta,c)$ are uniquely identifiable, and the inverse optimization estimator in \eqref{eq:inverse_obj} has a well-defined target.  
$\square$

\subsection{Proof of Proposition~\ref{prop:consistency} (Consistency)}
\label{proof:consistency}

\textbf{Proposition~\ref{prop:consistency}.}  
Suppose observed portfolios $\{x^t\}$ are generated from the forward problem 
\eqref{eq:forward_obj} with true parameters $(\theta^\ast,c^\ast)$, and that 
(i) the parameter space $\Theta$ is compact, 
(ii) $L_T(\theta)$ is continuous in $\theta$, and 
(iii) $\{(\mu^t,\Sigma^t)\}$ are i.i.d.\ draws with sufficient variation to ensure identifiability (Theorem~\ref{thm:identifiability}).  
Then the inverse estimator 
\[
\hat{\theta} = \arg\min_{\theta \in \Theta} L_T(\theta),
\quad L_T(\theta) = \frac{1}{T}\sum_{t=1}^T \|x^t - x^*(\theta)\|^2,
\]
is consistent: $\hat{\theta} \to_p \theta^\ast$ as $T \to \infty$.

\medskip
\textbf{Proof.}  
Define the population loss 
\[
L(\theta) = \mathbb{E}\big[\|x^t - x^*(\theta)\|^2\big],
\]
where the expectation is taken with respect to the distribution of $(\mu^t,\Sigma^t)$.  
By the uniform law of large numbers, if $\Theta$ is compact and $L_T(\theta)$ is continuous in $\theta$, then
\[
\sup_{\theta \in \Theta} |L_T(\theta) - L(\theta)| \to_p 0,
\]
so that the empirical loss converges uniformly to its population counterpart.  
By Theorem~\ref{thm:identifiability}, the population loss $L(\theta)$ is uniquely minimized at $\theta^\ast$, which ensures that $\theta^\ast$ is the well-defined target of estimation.  
Finally, by the argmin consistency theorem of Newey and McFadden (1994, Theorem 2.1), uniform convergence together with uniqueness of the minimizer implies that 
\[
\hat{\theta} \;\xrightarrow{p}\; \theta^\ast, \qquad T \to \infty.
\]
This establishes the consistency of the inverse estimator. $\square$

\subsection{Proof of Lemma~\ref{lem:robustness} (Robustness to Misspecification)}
\label{proof:robustness}

\textbf{Lemma~\ref{lem:robustness}.}  
Let the true transaction cost be $\phi(x)$, convex and homogeneous of degree $p \in [1,2]$, while the researcher estimates with a linear cost $c^\top x$.  
Define the approximation error
\[
\varepsilon = \sup_{x \in \mathcal{X}} \big| \phi(x) - c^\top x \big|.
\]
Then the inverse optimization estimator $(\hat{\theta},\hat{c})$ satisfies
\[
\|(\hat{\theta},\hat{c}) - (\theta^\ast,c^\ast)\| = O(\varepsilon).
\]

\medskip
\textbf{Proof.}  
The true forward objective is
\[
f_{\text{true}}(x;\theta) = \mu^\top x - \tfrac{\theta}{2} x^\top \Sigma x - \phi(x),
\]
while the approximated forward objective used in estimation is
\[
f_{\text{lin}}(x;\theta,c) = \mu^\top x - \tfrac{\theta}{2} x^\top \Sigma x - c^\top x.
\]
By definition of $\varepsilon$, the two objectives differ uniformly by at most $\varepsilon$ on the feasible set $\mathcal{X}$:
\[
| f_{\text{true}}(x;\theta) - f_{\text{lin}}(x;\theta,c) | \leq \varepsilon \quad \forall x \in \mathcal{X}.
\]
Because both are convex maximization problems over a compact set, their solutions are Lipschitz-stable with respect to perturbations in the objective (Bonnans and Shapiro, 2000, Ch.~4).  
Hence the optimizers $x^\ast_{\text{true}}$ and $x^\ast_{\text{lin}}$ satisfy
\[
\| x^\ast_{\text{true}} - x^\ast_{\text{lin}} \| = O(\varepsilon).
\]
Since the inverse estimator minimizes the squared deviation between observed portfolios (generated under $\phi$) and model-implied portfolios (generated under $c^\top x$), the parameter estimates $(\hat{\theta},\hat{c})$ inherit the same order of error.  
Therefore,
\[
\|(\hat{\theta},\hat{c}) - (\theta^\ast,c^\ast)\| = O(\varepsilon).
\]
This establishes the robustness claim. $\square$

\subsection{Proof of Theorem~\ref{thm:dyn_regret} (Dynamic Regret Bound)}
\label{proof:dyn_regret}

\textbf{Theorem~\ref{thm:dyn_regret}.}  
Assume (i) the forward objective $f(x;\theta)$ is convex in $x$ for each $\theta$,  
(ii) the mapping $\theta \mapsto x^*(\theta)$ is Lipschitz with constant $L_x$, and  
(iii) the preference sequence $\{\theta_t\}$ has bounded drift 
\[
\sum_{t=2}^T \|\theta_t - \theta_{t-1}\| \leq D.
\]  
Then the dynamic regret
\[
R_T = \sum_{t=1}^T f(x^*(\theta_t^{\text{true}})) - f(x^*(\hat{\theta}_t))
\]
satisfies
\[
R_T \leq O(\sqrt{T} + D).
\]

\medskip
\textbf{Proof.}  
If $\theta_t$ is constant, the problem reduces to online convex optimization with a fixed comparator $x^*(\theta)$, for which classical results (Zinkevich, 2003) yield $R_T = O(\sqrt{T})$.  
For varying $\theta_t$, regret can be decomposed into a static component relative to a fixed comparator and a drift component reflecting parameter changes.  
By Lipschitz continuity of $f(x^*(\theta))$ in $\theta$, each change $\|\theta_t - \theta_{t-1}\|$ contributes at most $O(\|\theta_t - \theta_{t-1}\|)$ to regret, so the cumulative drift term is bounded by $O(D)$.  
Combining the static $O(\sqrt{T})$ bound with the drift bound gives
\[
R_T \leq O(\sqrt{T} + D).
\]
This establishes the theorem. $\square$

\subsection{Proof of Corollary~\ref{cor:static_regret} (Static Preferences)}
\label{proof:static_regret}

\textbf{Corollary~\ref{cor:static_regret}.}  
If preferences are constant, i.e.\ $\theta_t = \theta^\ast$ for all $t$, then the drift term vanishes $(D=0)$.  
In this case, the dynamic regret bound simplifies to
\[
R_T = O(\sqrt{T}).
\]

\medskip
\textbf{Proof.}  
Setting $D=0$ in Theorem~\ref{thm:dyn_regret} removes the drift contribution, leaving only the static component $O(\sqrt{T})$.  
Hence, under static preferences, the regret matches the classical sublinear bound in online convex optimization. $\square$

\medskip
\textbf{Implication.}  
This corollary shows that our inverse estimator achieves sublinear regret in the classical static setting, fully consistent with standard OCO theory.  
It also confirms that Theorem~\ref{thm:dyn_regret} generalizes the well-known static case.

%---------------------------------
\section{Experimental Setup and Implementation Details}
\label{app:setup}

\subsection*{Computing Environment}
All simulations were executed on a workstation with the following configuration:
\begin{itemize}
    \item \textbf{Hardware:} NVIDIA RTX GPU, multi-core CPU, 128 GB RAM
    \item \textbf{Operating System:} Ubuntu 22.04.5 LTS under Windows Subsystem for Linux 2 (WSL2)
    \item \textbf{Software:} Python 3.10 (conda environment)
    \item \textbf{Libraries:} \texttt{cvxpy} 1.7.2 with \texttt{OSQP}, \texttt{numpy}, \texttt{scipy}
\end{itemize}

\subsection*{Simulation Scale}
\begin{itemize}
    \item Number of assets: $n = 10$
    \item Number of systematic factors: $k = 3$
    \item Time periods per trial: $T = 100$
    \item Monte Carlo replications: $R = 200$
    \item Total portfolio problems solved: $R \times T = 20{,}000$
\end{itemize}

\subsection*{Notes on Reproducibility}
\begin{itemize}
    \item Random seeds were fixed per replication. 
    \item All code was executed in Jupyter Notebook under WSL2.
    \item Source code and notebooks are available from the authors upon request.
\end{itemize}

\end{document}